\documentclass[aps,twocolumn,prd,superscriptaddress,showpacs,nofootinbib]{revtex4-1}

\usepackage[utf8]{inputenc}
\usepackage{diagbox}

\usepackage{mathtools}
\usepackage{amsfonts}
\usepackage{mathrsfs}
\usepackage{bbm}
\usepackage{slashed}

\usepackage{graphicx}
\usepackage{color}
\usepackage{array}

\usepackage{booktabs}
\usepackage{makecell}
\usepackage{epstopdf}
\usepackage[caption=false]{subfig}
\usepackage{multirow}

\usepackage{xspace}
\usepackage{siunitx}
\usepackage{hyperref}
\usepackage[nameinlink]{cleveref}
\usepackage{appendix}

\usepackage{xifthen}
\usepackage{xcolor}
\hypersetup{
	colorlinks,
	linkcolor={red!75!black},
	citecolor={blue!75!black},
	urlcolor={blue!75!black}
}

\newcommand{\td}{\ensuremath{\textrm{d}}}

\begin{document}

\title{Dyson-Schwinger equations towards cold-dense QCD matter with improved truncations}

\author{Zhan Bai}
\email{baizhan@siom.ac.cn}
\affiliation{State Key Laboratory of High Field Laser Physics and CAS Center for Excellence in Ultra-intense Laser Science, Shanghai Institute of
Optics and Fine Mechanics (SIOM), Chinese Academy of Sciences (CAS), Shanghai 201800, China}
\affiliation{Institute of Theoretical Physics, Chinese Academy of Sciences, Beijing 100190, China}

\author{Yu-Xin Liu}
\email{yxliu@pku.edu.cn}
\affiliation{Department of Physics and State Key Laboratory of Nuclear Physics and Technology, Peking University, Beijing 100871, China.}
\affiliation{Collaborative Innovation Center of Quantum Matter, Beijing 100871, China.}
\affiliation{Center for High Energy Physics, Peking University, Beijing 100871, China.}


\date{\today}

\begin{abstract}
We take the Dyson-Schwinger equation (DSE) approach of QCD to study the phase transition and the equation of state of cold dense matter.
Besides the bare vertex and Gauss gluon model, we take into account an improved truncation scheme, the CLRQ vertex and infrared-constant gluon model.
For the dynamical chiral symmetry breaking solution of the DSE,
we require that the emergence of quark number density to be at the chemical potential for the nuclear liquid-gas phase transition to take place,
by incorporating a chemical potential dependent modification factor to the {gluon model}.
The result shows that {our modified scheme} can not only describe the phase transition of the cold dense matter well but also the deduced equation of state of the matter can describe the recent astronomical observations consistently.
\end{abstract}

\maketitle

\section{Introduction}
\label{Sec:introduction}

Quantum chromodynamics (QCD) is believed to be the underlying theory of the strong interaction between quarks.
However, due to its complexity, the nature of some of its key features are still not fully understood.
For example, at low temperature and density, the quarks are always trapped in hadrons,
and the quarks inside hadrons would acquire a large mass.
These two features are known as confinement and dynamical chiral symmetry breaking (DCSB).
At high temperature or chemical potential, however,
the quarks escape from hadron, and their masses reduce to a small value.
This transition is known as hadron-quark phase transition,
and is essential in our understanding of the basic theory.

It is believed that at finite temperature and zero chemical potential,
the transition between the hadron and quark phases is a crossover.
In this region, the phase transition is understood well both theoretically and experimentally~\cite{Aoki:2006we,Aoki:2009sc,Borsanyi:2010bp,Bazavov:2011nk,Bhattacharya:2014ara,HotQCD:2014kol}.

At zero temperature and large chemical potential, however, the situation is unclear.
Although it is usually believed that the hadron-quark phase transition is in the first order,
there are still arguments that it should also be a crossover as the same as that in case of high temperature and zero chemical potential~\cite{Brandes:2021pti,Baym:2017whm,Fukushima:2020cmk}.
In this region, the lattice QCD approach fails because of the ``sign problem",
and one has to take advantage of the continuum approaches such as the Dyson-Schwinger equations (DSEs)~\cite{Roberts:1994dr,Roberts:2000aa,Maris:2003vk,Bashir:2012fs,Fischer:2018sdj,Gao:2020qsj,Gao:2020fbl,Gao:2021vsf}
and functional renormalization group (FRG)~\cite{Pawlowski:2005xe,Rosten:2010vm,Braun:2011pp,Pawlowski:2014aha,Dupuis:2020fhh} of QCD.

In particular, the DSE method is a non-perturbative, continuum approach of QCD.
It can simultaneously deal with the DCSB as well as the confinement,
and has been applied to many areas such as hadron properties and hadron-quark phase transition in different temperature and chemical potential region~\cite{Roberts:1994dr,Roberts:2000aa,Maris:2003vk,Bashir:2012fs,Fischer:2018sdj}.

The DSEs can be obtained by differentiating the generating functional of QCD,
and a complete set of DSEs contains infinite number of coupled equations (see, e.g., Ref.~\cite{Roberts:1994dr}).
For example, the DSE for the quark propagator, also known as ``gap equation'',
can be solved only after the dressed gluon propagator and dressed quark-gluon vertex are known (see Sec.~\ref{Sec::DSE} for detail).
The dressed vertex and gluon propagator can further be obtained by solving the higher order DSEs.
Therefore, in order to solve the DSEs, we have to do proper truncation.

The solutions of DSEs calculation certainly depend on the truncation schemes.
However, it is expected that with better and better truncation schemes we use,
the results should gradually converge to a fixed value,
and the remaining difference is an estimation of the uncertainty of the approach.
Therefore, for any DSE calculation,
it is essential to repeat the same procedure with more sophisticated truncation scheme and compare the results.
The calculation can be regarded reliable only after the results are convergent.

At the level of gap equation,
we can take models for the interaction vertex and the gluon propagator to do the truncation.
The most simple vertex is the bare vertex,
which is the leading order approximation, and is often referred to as ``rainbow-ladder (RL) approximation"~\cite{Munczek:1994zz,Bender:1996bb}.
And there are also vertices beyond the rainbow approximation,
such as the Ball-Chiu vertex~\cite{Ball:1980ax,Ball:1980ay} which satisfies the Ward-Takahashi identity,
the Chang-Liu-Roberts-Qin (CLRQ) vertex~\cite{Chang:2010hb,Chang:2011ei,Qin:2013mta} which includes the dynamical chiral symmetry breaking effect and explains the anomalous magnetic moments,
Tang-Gao-Liu (TGL) vertex~\cite{Tang:2019zbk} which includes all the Lorentz structures,
and so on.

As for the gluon model,
there are contact model~\cite{Roberts:2011wy} which assumes that the interaction appears only at infinitesimal distance,
and Munczek model~\cite{Munczek:1983dx},which assumes that the interaction is independent of the separation.
There are more realistic models such as the Maris-Tandy model~\cite{Maris:1999nt} in which the interaction vanishes at infrared,
and Qin-Chang (QC) model~\cite{Qin:2011dd}, whose interaction remains a constant value at infrared.

Among theses different truncation schemes,
the combination of CLRQ vertex and QC gluon model is found to bridge very well the bottom-up scheme with the {\it ab initio} computation in continuum QCD~\cite{Binosi:2014aea},
and have been applied to study the QCD phase transition at finite temperature and chemical potential~\cite{Gao:2016qkh}.

The DSE approach has also been taken to study the cold dense matter~\cite{Chen:2008zr,Chen:2011my,Chen:2012zx,Muller:2013pya,Chen:2015mda,Chen:2016ran,Xu:2015jwa,Bai:2017wvk,Bai:2021non,Qin:2023zrf,Bai:2019jtl},
which is relevant in neutron stars.
In Refs.~\cite{Chen:2011my,Chen:2012zx,Chen:2015mda,Chen:2016ran,Bai:2017wvk},
the gap equation was solved by assuming a damping factor in the gluon model, which simulates the effect of asymptotic freedom at large chemical potential.
In these works, there remains a free parameter which controls the speed for the quark matter to approach asymptotic freedom.

In Ref.~\cite{Bai:2021non}, we once attempted to fix this free parameter.
It has been confirmed that, at baryon chemical potential $\mu_{B}=923\;$MeV,
   which is the proton mass minus the binding energy,
   the matters will appear from the vacuum through a first-order phase transition.
This is referred to as the ``nuclear liquid-gas phase transition"~\cite{Pochodzalla:1995xy,Chomaz:2001ngp,DAgostino:2005qpq,Drews:2016wpi} (we take this terminology in this paper).
In DSE approach, there also exist a critical chemical potential where the quark number density of the hadron matter becomes non-zero.
Therefore, by requiring this critical chemical potential to be the same as that for nuclear liquid-gas phase transition,
we fix the parameter in the damping factor.

As we have stated, the convergence of the DSE result should be checked.
Therefore, in this paper, based on our previous work,
we will take the DSE with improved truncation schemes to study the hadron-quark phase transition.
The new truncation schemes include QC gluon and CLRQ vertex.
We will also show that a modification for the truncation is necessary in order to reduce the uncertainty of the DSEs.

Apart from these theoretical studies, verifications from experiments are also essential in our understanding of the phase transition.
However, it is now impossible to create the matter with high density and zero temperature on earth,
and one has to take astronomical observations of neutron stars to study the properties of the matter in this region.

It has been well known that neutron star is one of the most compact objects in the universe,
and it is very likely that the inner core of neutron star can be so dense that the hadron-quark phase transition takes place.
The most useful information of neutron star that can be observed on earth is their mass.
The large-mass neutron star observed in recent years requires that the equation of state (EOS) of the dense matter should be stiff~\cite{Demorest:2010bx,Antoniadis:2013pzd,Fonseca:2016tux,NANOGrav:2017wvv,NANOGrav:2019jur,Linares:2018ppq}.
Also, the detection of the gravitational wave suggests that the neutron star has small tidal deformability, and the EOS should be soft~\cite{LIGOScientific:2017zic,LIGOScientific:2018cki,Margalit:2017dij,Rezzolla:2017aly,Shibata:2019ctb,Ruiz:2017due,Shibata:2017xdx,Annala:2017llu,Annala:2019puf}.
These two kinds of observations together provide constraints on the theoretical models of the neutron star matter.
The joint mass-radius observation of the neutron star from the Neutron Star Interior Composition Explorer (NICER) mission
provides significant information for the profile of the neutron star~\cite{Riley:2019yda,Miller:2019cac,Riley:2021pdl,Miller:2021qha},
which is also essential in the study of hadron-quark phase transition.

This paper is organized as follows:
after this introduction, we will briefly describe the DSE approach in Sec.~\ref{Sec::DSE}.
Then, in Sec.~\ref{Sec::Modification}, we will prove the necessity of modifying the current truncation schemes,
and then fix the parameter of our modification with chemical potential for the liquid-gas phase transition to take place.
In Sec.~\ref{Sec::PhaseTransition}, we will make use of the modified truncation schemes to study the hadron-quark phase transition.
We will also construct the EOS for the matter involving the phase transition
and calculate the neutron star mass-radius relation in Sec.~\ref{Sec::NeutronStar}.
Finally, in Sec.~\ref{Sec::summary}, we give a brief summary and some remarks.

\section{Dyson-Schwinger equation approach}\label{Sec::DSE}

In this section, we describe the DSE approach at zero temperature but finite chemical potential.

\begin{figure}[!htb]
\includegraphics[width=0.45\textwidth]{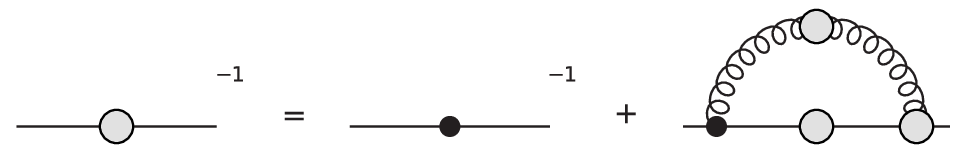}
\caption{
The diagrammatic representation for the Dyson-Schwinger equation of the quark propagator.
The solid line with gray circle denotes the dressed quark propagator,
the solid line with black circle denotes the bare quark propagator,
the curly line with gray circle denotes the dressed gluon propagator,
and the black and gray circles stands for the bare vertex and the dressed vertex, respectively.
}
\label{fig:gap_equation}
\end{figure}

The starting point is the gap equation in QCD, {\it i.e.}, the DSE for the quark propagator.
A diagrammatic representation of the equation is shown in Fig.~\ref{fig:gap_equation}.
Written explicitly, it reads:
\begin{equation}
S(p;\mu)^{-1}=Z_{2}^{} [\textrm{i} \vec{\gamma} \cdot  \vec{p} + \textrm{i} \gamma_{4}^{} (p_{4}^{} + \textrm{i}\mu) + m_{q}^{}] + \Sigma(p;\mu),
\end{equation}
where $S(p;\mu)$ is the quark propagator, $\Sigma(p;\mu)$ is the renormalized self-energy of the quark:
\begin{equation}
\begin{split}
\Sigma(p;\mu)=& \, Z_{1}^{} \int^{\Lambda}\frac{\textrm{d}^{4} q}{(2\pi)^{4}}
          g^{2}(\mu)D_{\rho\sigma}^{} (p-q;\mu)\\
		&\; \times\frac{\lambda^{a}}{2} \gamma_{\rho} S(q;\mu) \Gamma_{\sigma}^{a}(q,p;\mu),
\end{split}
\end{equation}
where $\int^{\Lambda}$ is the translationally regularized integral,
$\Lambda$ is the regularization mass-scale.
$g(\mu)$ is the strength of the coupling, $D_{\rho\sigma}^{}$ is the dressed gluon propagator,
$\Gamma_{\sigma}^{a}$ is the dressed quark-gluon vertex,
$\lambda^{a}$ is the Gell-Mann matrix, and $m_{q}^{}$ is the current mass of the quark.
$Z_{1,2}^{}$ are the renormalization constants.
In this paper, we will apply truncation schemes with which the ultraviolet integration is finite,
so we can avoid doing renormalization and take $Z_{1,2}^{}=1$.

At finite chemical potential,
the quark propagator can be decomposed according to the Lorentz structure as:
\begin{equation}\label{eqn:deomposition}
\begin{split}
S(p;\mu)^{-1}=&\; \textrm{i}\vec{\gamma} \cdot  \vec{p} A(|\vec{p}|, p_{4}; \mu)
+ B(|\vec{p}|, p_{4}; \mu)\\
		& + \textrm{i} \gamma_{4}^{} (p_{4} + \textrm{i} \mu) C(|\vec{p}|, p_{4}; \mu) \, ,
\end{split}
\end{equation}
 where $A$, $B$ and $C$ are scalar functions.
A complete decomposition should include another term proportional to $\sigma_{\mu\nu}$,
  but this term contributes little, and is usually omitted~\cite{Roberts:2000aa,Fischer:2018sdj}.

At zero chemical potential, a commonly used ansatz for the dressed gluon propagator and the dressed quark-gluon interaction vertex is:
\begin{equation}
\begin{split}
Z_{1}^{} g^{2}  &D_{\rho\sigma}^{}(p- q)\Gamma_{\sigma}^{a}(q,p)\\
		=&\mathcal{G}\left((p - q)^{2} \right)D_{\rho\sigma}^{\textrm{free}}(p-q) 
    \times\frac{\lambda^{a}}{2} \Gamma_{\sigma}^{}(q,p) \, ,
\end{split}
\end{equation}
where
\begin{equation}
D_{\rho\sigma}^{\textrm{free}}\left(k\equiv p-q\right)=\frac{1}{k^{2}} \Big( \delta_{\rho\sigma}^{} - \frac{k_{\rho}^{} k_{\sigma}^{}}{k^{2}} \Big) \, .
\end{equation}
$\mathcal{G}(k^{2})$ is the effective interaction to be introduced in a model,
and $\Gamma_{\sigma}^{}$ is the quark-gluon vertex.

The simplest truncation for the quark-gluon vertex is the bare vertex, or rainbow-ladder (RL) truncation:
\begin{equation}\label{eqn:bare_vertex}
\Gamma_{\sigma}^{}(q,p) = \gamma_{\sigma}^{} \, .
\end{equation}

In this paper, 
in order to check the convergence and uncertainty of our calculation,
we also adopt an improved vertex, the CLRQ vertex:
\begin{equation}\label{eqn:Gamma_CLR}
\Gamma^{\rm CLRQ}_{\mu}=\Gamma_{\mu}^{\rm BC}+\Gamma_{\mu}^{\rm ACM},
\end{equation}
where $\Gamma_{\mu}^{\rm BC}$ is the Ball-Chiu vertex.
At zero temperature, it reads
\begin{equation}
\begin{split}
\Gamma_{\mu}^{\rm BC}&(\tilde{q},\tilde{p})\\
=&\; \gamma_{\mu}^{T}\Sigma_{A}+\gamma_{\mu}^{L}\Sigma_{C}+(\tilde{p}+\tilde{q})_{\mu}\Big{[} \frac{1}{2}\gamma_{\alpha}^{T}(\tilde{p}+\tilde{q})_\alpha\Delta_{A}\\
	&\; +\frac{1}{2}\gamma_{\alpha}^{L}(\tilde{p}+\tilde{q})_{\alpha}\Delta_{C}-i\Delta_{B}\Big{]},
\end{split}
\end{equation}
with
\begin{equation}
\tilde{p}=(\vec{p},\tilde{p}_{4}),\qquad \tilde{q}=(\vec{q},\tilde{q}_{4}),
\end{equation}
\begin{equation}
\tilde{p}_{4} = p_{4} + \textrm{i} \mu_{q},\qquad \tilde{q}_{4} = q_{4} + \textrm{i} \mu_{q},
\end{equation}
\begin{equation}\label{eqn:SigmaF_DeltaF}
\begin{split}
\Sigma_{F}&=\frac{1}{2}\left[F(|\vec{q}|,{q}_{4};\mu_{q})+F(|\vec{p}|,{p}_{4};\mu_{q})\right],\\
\Delta_{F}&=\frac{F(|\vec{q}|,{q}_{4};\mu_{q})-F(|\vec{p}|,{p}_{4};\mu_{q})}{\tilde{q}^2-\tilde{p}^2},
\end{split}
\end{equation}
where $F=A,B,C$ and $\gamma_{\mu}^{T}=\gamma_{\mu}-\gamma_{\mu}^{L}$, $\gamma_{\mu}^{L}=u_{\mu}\gamma_{\alpha}u_{\alpha}$,
$u=(0,0,0,1)$.

The $\Gamma_{\mu}^{\rm ACM}$ in Eq.(\ref{eqn:Gamma_CLR}) is the transverse part in the vertex that characterizes the DCSB effect in the quark-gluon vertex
through the anomalous chromomagnetic moments, which reads
\begin{equation}
\Gamma_{\mu}^{\rm ACM}=\Gamma_{\mu}^{{\rm ACM}_4}+\Gamma_{\mu}^{{\rm ACM}_5},
\end{equation}
\begin{equation}
\Gamma_{\mu}^{{\rm ACM}_4}=\left[T_{\mu\nu}l_{\nu}\gamma\cdot k+\textrm{i}T_{\mu\nu}\gamma_{\nu}\sigma_{\rho\sigma}l_{\rho}k_{\sigma}\right]\tau_{4}(\tilde{p},\tilde{q}),
\end{equation}
\begin{equation}
\Gamma_{\mu}^{{\rm ACM}_{5}}=\sigma_{\mu\nu}k_{\nu}\tau_{5}(\tilde{p},\tilde{q}),
\end{equation}
\begin{equation}
\tau_{4}=\frac{2\tau_{5}(\tilde{p},\tilde{q})\left[2(M(\tilde{p})+M(\tilde{q}))\right]}{\tilde{p}^{2}+M(\tilde{p})^2+\tilde{q}^{2}+M(\tilde{q})^2},
\end{equation}
\begin{equation}
\tau_{5}=\eta\Delta_{B},
\end{equation}
where $\sigma_{\mu\nu}=\frac{i}{2}[\gamma_{\mu},\gamma_{\nu}]$,
$T_{\mu\nu}=\delta_{\mu\nu}-k_{\mu}k_{\nu}/k^2$,
$k_{\mu}=(\tilde{p}-\tilde{q})_{\mu}$,
	$l_{\mu}=\frac{(\tilde{p}+\tilde{q})_{\mu}}{2}$,
$M(x)=B(x)/A(x)$, $\Delta_{B}$ is defined in Eq.~(\ref{eqn:SigmaF_DeltaF}) and $\eta$ is a parameter.
In this paper, we take $\eta=0.65$ as that  in Ref.~\cite{Gao:2016qkh}.

For the effective interaction ${\mathcal{G}}(k^{2})$, one of the widely used model is the Maris-Tandy (MT) model~\cite{Maris:2003vk,Holl:2005vu,Bhagwat:2006xi,Eichmann:2008ef,Nguyen:2011jy,Eichmann:2011vu}:
\begin{equation}\label{eqn:MT}
\frac{\mathcal{G}(k^{2})}{k^{2}}=\frac{4\pi^{2} D}{\omega^{6}}k^{2} \textrm{e}^{-k^{2}/\omega^{2}} +\alpha_{\rm pQCD}\, .
\end{equation}

As can be seen from Eq.~(\ref{eqn:MT}), the interaction of MT model vanishes in the infrared domain ($k^{2}\approx 0$).
However, the modern DS equation and lattice QCD studies indicate that the gluon propagator is a bounded,
regular function of spacelike momenta, which achieves its maximum value at $k^{2}= 0$~\cite{Bogolubsky:2009dc,Boucaud:2010gr,Oliveira:2010xc,Cucchieri:2011ig,Aguilar:2012rz,Ayala:2012pb,Dudal:2012zx,Strauss:2012dg,Zwanziger:2012xg,Blossier:2013te}.
Therefore, the Qin-Chang (QC) model, which has a non-vanishing infrared interaction,
is more sophisticated~\cite{Qin:2011dd,Chang:2011ei,Binosi:2014aea}:
\begin{equation}\label{eqn:QC}
\frac{\mathcal{G}(k^{2})}{k^{2}}=\frac{8\pi^{2} D}{\omega^{4}} \textrm{e}^{-k^{2}/\omega^{2}} +\alpha_{\rm pQCD} \, .
\end{equation}

The $\alpha_{\rm pQCD}$ in Eqs.~(\ref{eqn:MT}) and (\ref{eqn:QC}) is the ultraviolet perturbative term.
In zero temperature and large chemical potential, the quark properties are dominated by the infrared behavior.
Therefore, we would omit the $\alpha_{\rm pQCD}$ term for better numerical behavior.
To distinguish from the original model,
   we will use ``Gauss model'' to refer to MT model without perturbative term,
   and use ``infrared-constant (IC) model'' to refer to QC model without perturbative term.

The $D$ and $\omega$ in Eqs.~(\ref{eqn:MT}) and (\ref{eqn:QC}) are parameters of the models.
Practical calculations have shown that the properties of hadron is insensitive to the value of $\omega$ for a fixed value of$(D\omega)$,
    when $\omega$ is in the range $[0.4,0.6]\;$GeV (see, e.g., Refs.~\cite{Maris:1999nt,Bashir:2012fs}),
and we fix $\omega=0.5\;$GeV in this paper.
The value of $D$ and current mass for $u$ and $d$ quark is fixed by fitting the pion mass and decay constant,
    and the strange quark mass is determined by fitting the kaon mass or directly using the value from particle data group.
In this paper, we consider three combinations of the vertex and gluon models,
   and the value of the parameters for different combinations are shown in Table~\ref{Tab:parameter}.

\begin{table}[!htp]
\caption{The vertex and gluon models we use and corresponding parameters.}\label{Tab:parameter}
\begin{tabular}{|c|c|c|c|c|c|}
\hline
&vertex&gluon&$D({\textrm{GeV}^2})$&$m_{u,d}$(MeV)&$m_{s}$(MeV)\\
\hline
DSE1 & RL & Gauss & 1.0~\cite{Chang:2009zb}  & 5~\cite{Chang:2009zb}  & 115~\cite{Alkofer:2002bp} \\
DSE2 & RL & IC & 1.024~\cite{Qin:2011xq}& 3.4~\cite{Qin:2011xq}  & 82~\cite{Qin:2011xq} \\
DSE3 & CLRQ & IC & 0.2812~\cite{Chang:2011ei} & 3.7~\cite{Chang:2011ei}  & 95~\cite{ParticleDataGroup:2018ovx} \\
\hline
\end{tabular}
\end{table}

We mention that, the DSE1, DSE2 and DSE3 truncation schemes are each more sophisticated than the previous one,
and we would naively expect that the results will gradually converge as the scheme gets more improved.
However, as we will soon see,
there will be no convergence unless a modification to the truncation schemes is introduced.

\section{Modification of the Truncation Schemes}\label{Sec::Modification}

\subsection{Emergence of quark number and comparison of the schemes}

After determining the truncation scheme, we can solve the DSEs numerically.

As we have mentioned, there are two solutions for the gap equation.
The one with large mass function $M(p)=B(p)/A(p)$ is known as the Nambu solution,
and corresponds to the DCSB--confined phase, or the hadron phase.
The other is the Wigner solution, with small mass function,
which corresponds to the DCS--unconfined phase, or the quark phase.
For hadron matter in low chemical potential region,
there exists a ``silver blaze'' property,
which means that the number density remains zero until a critical chemical potential.
Such a chemical potential should correspond to the nuclear liquid-gas phase transition.

It is remarkable that the liquid-gas phase transition we referred here is understood as~\cite{Drews:2016wpi}:
at the critical chemical potential, $\mu_{B,c}(T)$,
the potential energy of the system has two minima.
In the $T=0$ limit, one minimum corresponds to the phase where the baryon number density is zero,
and the other corresponds to the phase where the baryon number density is nonzero.
The phase transition at $\mu_{B,c}^{}$ is of first order.
For detailed discussion of such a concept, please see Ref.~\cite{Drews:2016wpi}.

The value of the critical baryon chemical potential is the mass of proton minus the binding energy,
and takes $\mu_{B,c}^{}=923\,$MeV.
The corresponding quark chemical potential is then $\mu_{q,c}=\mu_{B,c}/3\approx 0.3077\;$GeV.

In the framework of DSE, we can calculate the quark number density from our numerical solution.
After solving the gap equation at some quark chemical potential,
the number density of quarks can be obtained through~\cite{Chen:2008zr}:
\begin{equation}\label{eqn:nq}
n_{q}^{}(\mu_{q}) = 6\int\frac{\textrm{d}^{3}p}{(2\pi)^{3}} f_{q}^{}(|\vec{p}|;\mu_{q}),
\end{equation}
where $f_{q}^{}$ is the distribution function and reads
\begin{equation}\label{eqn:distribution}
\begin{split}
f_{q}^{}(|\vec{p}|;\mu_{q}) =& \; \frac{1}{4\pi}\int_{-\infty}^{\infty}\textrm{d}p_{4}^{} \textrm{tr}_{\textrm{D}}^{}[-\gamma_{4}^{} S_{q}^{}(p;\mu_{q})] \\
	=& \; \frac{1}{\pi}\int_{-\infty+i\mu_{q}}^{\infty+i\mu_{q}}\td \tilde{p}_{4}\frac{i\tilde{p}_4C(\vec{p}^2,\tilde{p}_{4}^2)}{\mathcal{M}},
\end{split}
\end{equation}
where the trace is for the spinor indices,
	and the denominator is:
\begin{equation}\label{eqn:denominator}
\mathcal{M}=\vec{p}^{2}A^2(|\vec{p}|^2,\tilde{p}_{4}^2)+\tilde{p}_{4}^{2}C^2(|\vec{p}|^2,\tilde{p}_{4}^2)+B^2(|\vec{p}|^2,\tilde{p}_{4}^2).
\end{equation}

In principle, we can calculate the number density from Eq.(\ref{eqn:nq}),
and find the chemical potential at which the number density turns from zero to non-zero.
However, because of the numerical error, the number density has a small but non-zero value even in vacuum,
and it is hard to directly find the critical chemical potential.
Therefore, we instead study the pole structure of the integrand in Eq.~(\ref{eqn:distribution}),
and determine the critical chemical potential more accurately,
as has been done in Ref.~\cite{Bai:2021non}.

\begin{figure}[!htbp]
\includegraphics[width=0.43\textwidth]{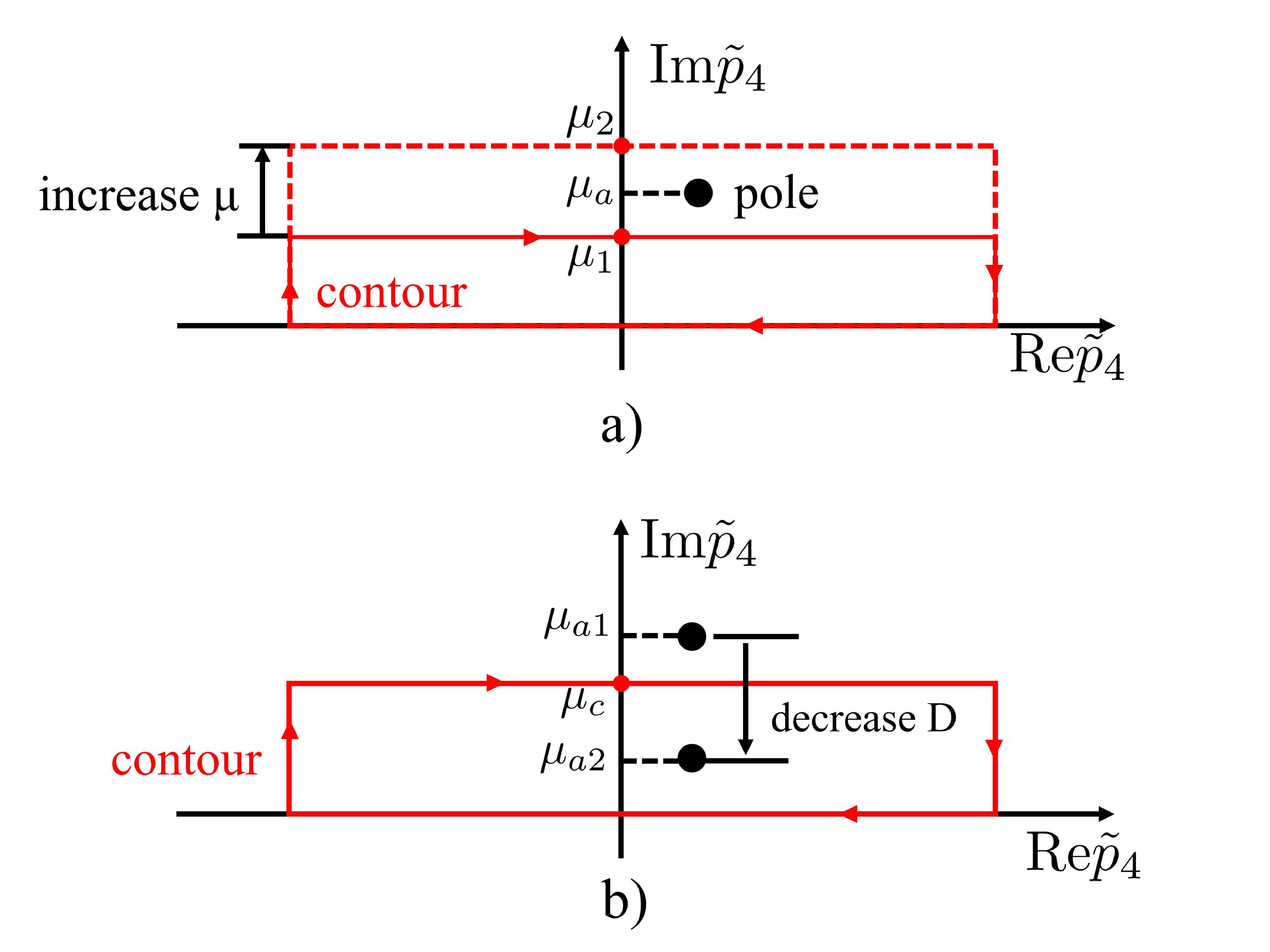}
\vspace*{-4mm}
\caption{(color online) Schematic feature of the contour and the singularities for the integration in Eq.~(\ref{eqn:distribution})\ 
(Taken from Ref.~\cite{Bai:2021non}).
Upper panel: by increasing the chemical potential, the contour may include the pole, and the number density becomes nonzero.
Lower panel: for a fixed chemical potential, by decreasing the coupling constant $D$,
the imaginary part of the pole becomes small and moves inside the contour,
and the number density becomes nonzero.
}
\label{fig:singularity}
\end{figure}

From Eq.(\ref{eqn:denominator}), we see that the denominator $\mathcal{M}$ has zeros at
\begin{equation}
	\tilde{p}_{4}=\pm\frac{\textrm{i}\sqrt{\vec{p}^{2}A^2+B^2}}{C},
\end{equation}
which correspond to poles in Eq.~(\ref{eqn:distribution}).
Since for Nambu solution, the function $B(|\vec{p}|^2,\tilde{p}_{4}^{2})$ has a large value,
and both $A(|\vec{p}|^{2},\tilde{p}_{4}^{2})$ and $C(|\vec{p}|^{2},\tilde{p}_{4}^{2})$ is of order 1,
the pole of $\tilde{p}_{4}$ has a large imaginary part.

As is shown in the upper panel of Fig.~\ref{fig:singularity},
the integration in Eq.~(\ref{eqn:distribution}) remains zero if $\mu_{q}$ is small and there is no singularity inside the contour $(-\infty + \textrm{i} \mu_{q}) \rightarrow (\infty + \textrm{i}\mu_{q}) \rightarrow (\infty) \rightarrow(-\infty) \rightarrow (-\infty + \textrm{i}\mu_{q})$
(we use $\mu_{a}$ to denote the imaginary part of the pole).
However, if we increase $\mu_{q}$, the contour will finally include the pole,
and the number density becomes nonzero, which is shown as the dashed lines in the upper panel in Fig.~\ref{fig:singularity}.
The critical chemical potential corresponds to the contour passing right through the pole.

This pole corresponds to a zero point of $\mathcal{M}(|\vec{p}|,\tilde{p}_4)$.
Therefore, we can scan $(|\vec{p}|,\tilde{p}_4)$,
and find the maximum value of $1/|\mathcal{M}|$.
At critical chemical potential, we should have $\textrm{Max}[1/|\mathcal{M}|]$ divergent.

\begin{figure}[!htbp]
\includegraphics[width=0.43\textwidth]{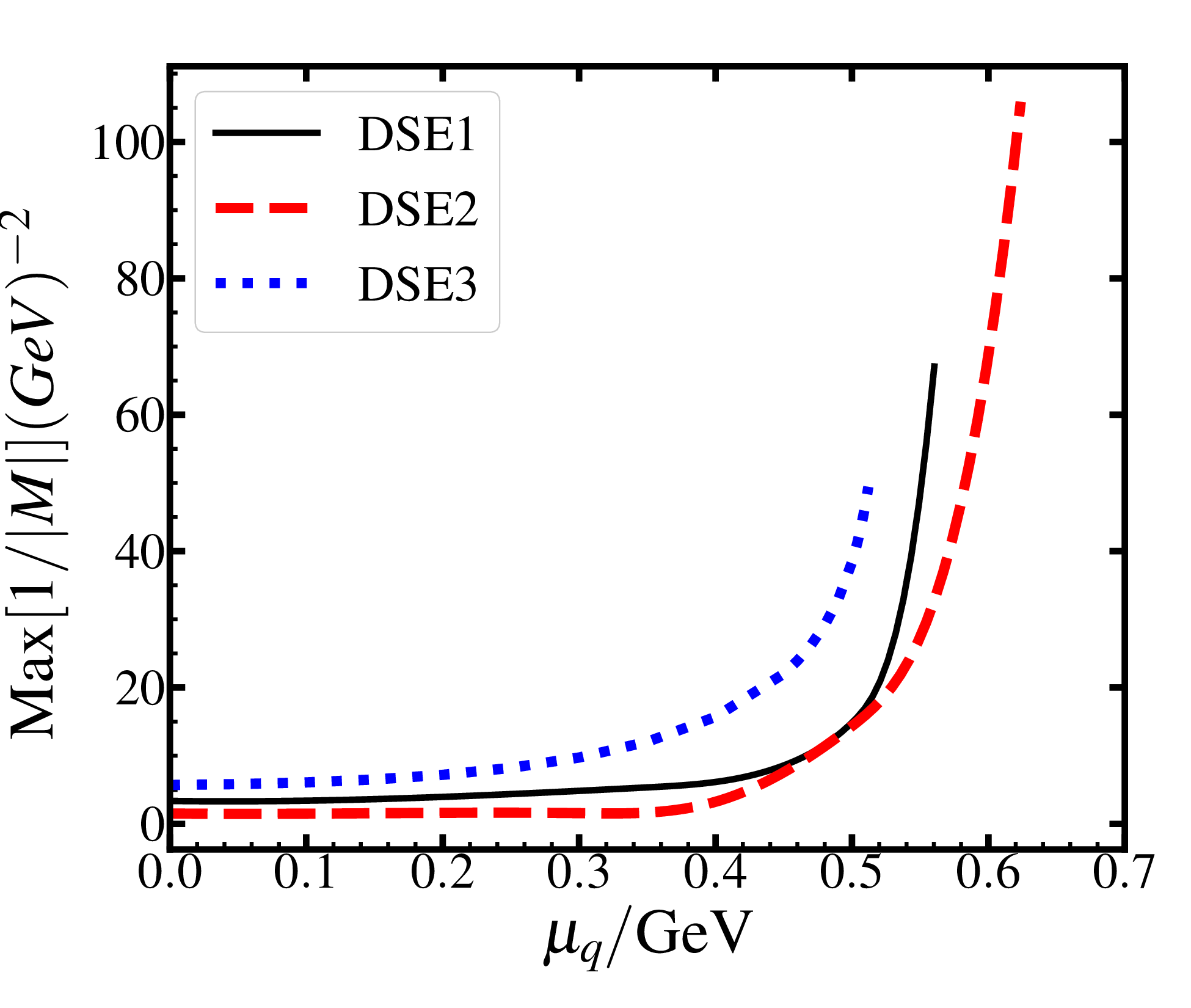}
\vspace*{-5mm}
\caption{(color online)
The obtained maximum value of $1/|\mathcal{M}|$ as a function of the quark chemical potential.
The black solid line corresponds to the result using the RL vertex and Gauss gluon model,
the red dashed line corresponds to the result using the RL vertex and IC gluon model,
and the blue dotted line corresponds to the result using the CLRQ vertex and IC gluon model.
}
\label{fig:convergence}
\end{figure}

In Fig.~\ref{fig:convergence}, we show the obtained variation behavior of $\textrm{Max}[1/|\mathcal{M}|]$ as a function of quark chemical potential.
As can be seen from the figure, for DSE1, DSE2 and DSE3 truncation scheme,
the critical chemical potential $\mu_{q,c}=0.566$, $0.630$ and $0.512\;$GeV, respectively.

These calculated results are obviously problematic.
This is primarily reflected in the following two points:

Firstly, the critical chemical potentials correspond to the emergence of hadron matter,
and the system will remain vacuum below the $\mu_{q,c}$.
As we have stated, the hadron matter should appear at liquid-gas phase transition,
which occurs at $\mu_{q,c}\approx 0.3077\;$GeV,
i.e., our directly calculated result is in great contradiction with the data in real world.

Secondly, it is expected that with better and better truncation schemes taken,
the calculated results of the DSEs should gradually approach to a fixed value,
with only very small dependence on the higher order diagrams,
i.e., the uncertainty of DSEs gradually reduces.
However, the results in Fig.~\ref{fig:convergence} show no convergence at all.

Therefore, in order to study QCD matter with zero temperature and high chemical potential with the DSE approach,
we must modify our truncation scheme.
In Refs.~\cite{Chen:2011my,Chen:2012zx,Chen:2015mda,Chen:2016ran,Bai:2017wvk,Bai:2021non},
a chemical potential dependent truncation scheme has been introduced.
In the following, we will reiterate the corresponding modification scheme,
and determine the extra parameter with the nuclear liquid-gas phase transition.

\subsection{Modification to truncation scheme}

In order to correctly reproduce the critical chemical potential,
we need to modify our truncation scheme so that the pole of Eq.~(\ref{eqn:distribution}) enters the contour at $\mu_{q,c}=0.3077\;$GeV.
As shown in the lower panel of Fig.~\ref{fig:singularity},
we can adjust the position of the pole by tuning the value of coupling constant $D$,
which is defined in Eqs.~(\ref{eqn:MT}) and (\ref{eqn:QC}).

\begin{figure}[!htb]
\includegraphics[width=0.40\textwidth]{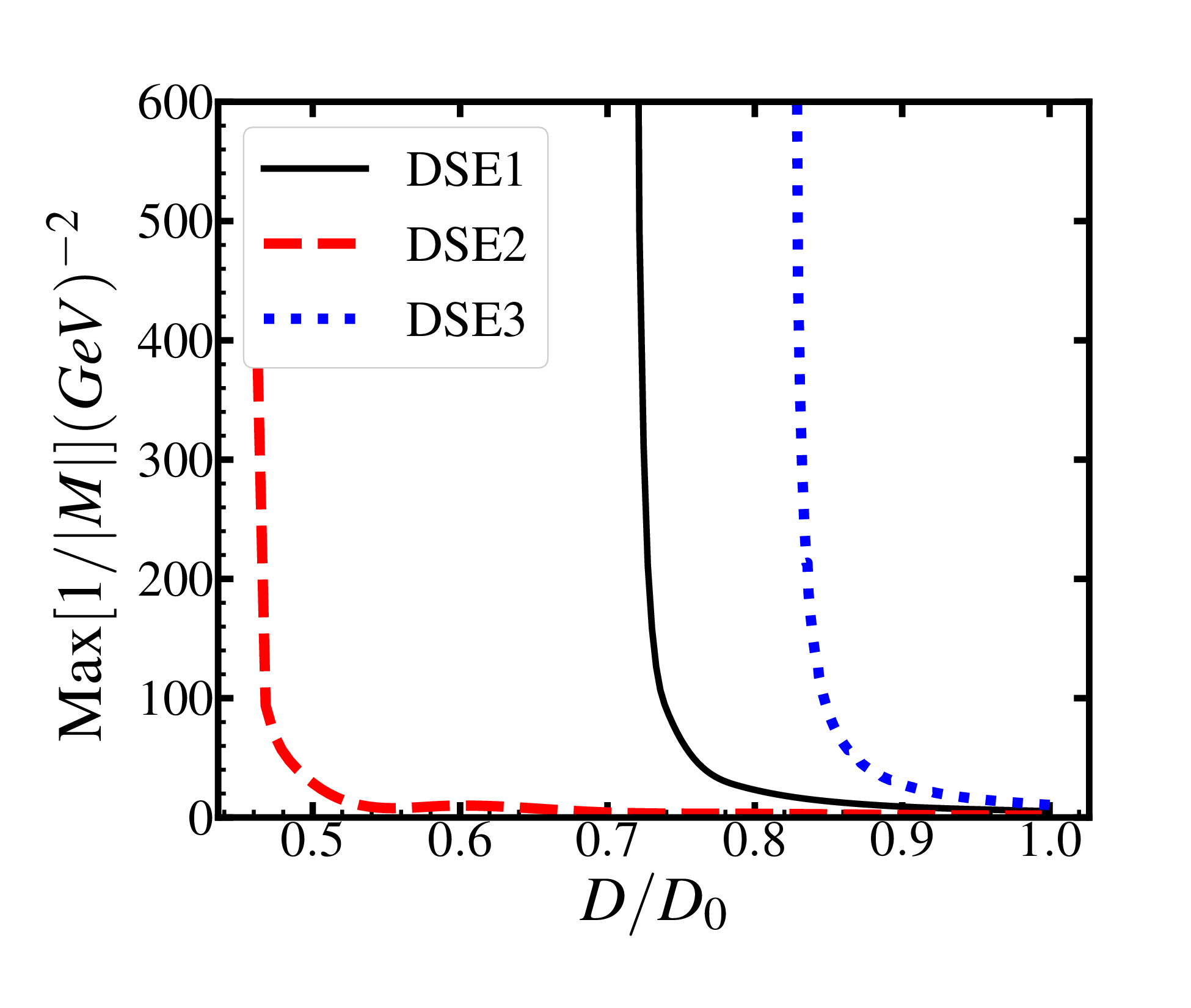}
\vspace*{-3mm}
\caption{(color online) The calculated relation between the $\textrm{Max}\left(1/|\mathcal{M}|\right)$ and the coupling strength $D$.
	$\mathcal{M}$ is defined in Eq.~(\ref{eqn:denominator}),
		and $D_{0}$ is the value displayed in Table~\ref{Tab:parameter}.
The black solid line corresponds to the result using the RL vertex and Gauss gluon model,
the red dashed line corresponds to the result using the RL vertex and IC gluon model,
and the blue dotted line corresponds to the result using the CLRQ vertex and IC gluon model.
}
\label{fig:pole}
\end{figure}

Therefore, we fix the chemical potential at $\mu_{q}=\mu_{q,c}$,
and gradually reduce the $D$ to look for when $\textrm{Max}[1/|\mathcal{M}|]$ is divergent,
i.e., the singularity enters the contour.
The result is shown in Fig.~\ref{fig:pole}.
From the figure, we can notice that the $1/|\mathcal{M}|$ diverge at $D_{c}/D_{0}=0.716, \, 0.460,$ and $0.826$ for DSE1, DSE2, DSE3 model, respectively,
where $D_{0}$ is the value listed in Table~\ref{Tab:parameter}.

In Refs.~\cite{Chen:2011my,Chen:2012zx,Chen:2015mda,Chen:2016ran,Bai:2017wvk,Bai:2021non}
a chemical potential dependence was introduced for the coupling strength $D$
(which is defined in Eqs.~(\ref{eqn:MT}) and (\ref{eqn:QC})):
\begin{equation}\label{eqn:D_mu}
D(\mu_{q})=D_{0}h(\mu_{q}).
\end{equation}

Therefore, the modification in Eq.~(\ref{eqn:D_mu}) must satisfy three constraints:
\begin{enumerate}
	\item At $\mu_{q}=0$, we should have $h(\mu_{q})=1$, since the coupling strength $D$ is obtained by fitting the hadron properties in vacuum.
	\item At $\mu_{q}=\infty$, we should have $h(\mu_{q})=0$, in order to approach the asymptotic freedom.
	\item At $\mu_{q}={\mu_{B,c}^{}}/3$, we should have $h(\mu_{q})=0.716$, $0.460$ and $0.826$ for the DSE1, DSE2, DSE3 scheme, respectively,
	in order to reproduce the nuclear liquid-gas phase transition chemical potential.
\end{enumerate}

For the modification function $h(\mu_{q})$, we follow our previous work~\cite{Bai:2021non}:
\begin{equation}\label{eqn:damping}
h(\mu_{q})=\left(1+\frac{\mu_{q}^2}{\mu_{q,c}^2}\right)\exp(-\beta\mu_{q}^2/\omega^2),
\end{equation}
where $\mu_{q,c}=923/3\;$MeV is the quark chemical potential corresponding to the nuclear liquid-gas phase transition,
$\beta$ is a parameter to be fixed.
And by fitting the value of $h(\mu_{q,c})$,
we have $\beta_{1}=2.714$, $\beta_{2}=3.861$ and $\beta_{3}=2.337$ for DSE1, DSE2 and DSE3 truncation scheme, respectively.

Therefore, by introducing the modification factor,
all the three sets of truncation schemes successfully recovers the nuclear liquid-gas chemical potential.

Also, since the modification factor damps exponentially at large chemical potential,
the quark propagator will automatically approach to that of the asymptotic freedom at large density,
which recovers the key property of QCD.

In the following, we will make use of these modified truncation schemes to study the hadron-quark phase transition
as well as the EOS of the neutron star matter.

\section{Phase transition region}\label{Sec::PhaseTransition}

\subsection{Isosymmetric phase transition and coexistence region}\label{sec::isosymmetric}

\begin{figure}[!htb]
\includegraphics[width=0.45\textwidth]{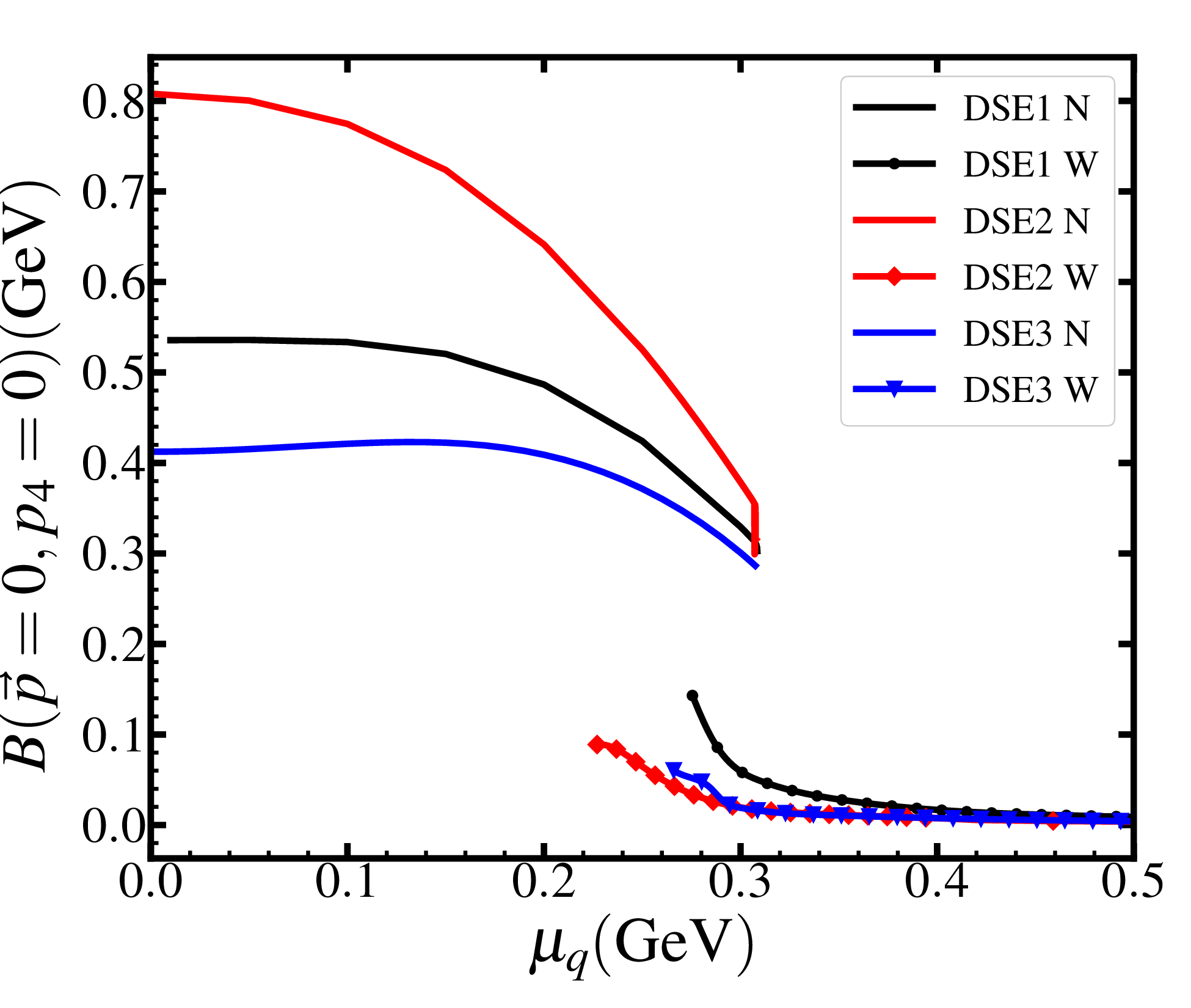}
\vspace*{-4mm}
\caption{(color online) The calculated $B(p^{2}=0)$ as a function of quark chemical potential in case of different vertex and gluon model and different solutions.
The solid lines correspond to Nambu solution, the solid lines with symbols corresponds to Wigner solution.
The black lines correspond to the solutions  with the RL vertex and Gauss gluon model,
the red lines correspond to the solutions with RL vertex and IC gluon model,
and the blue lines correspond to the solutions with CLRQ vertex and IC gluon model.
}
\label{fig:mu_B0}
\end{figure}

The Nambu solution of the DSE corresponds to the DCSB--confined phase (i.e., the hadron matter),
    and the quarks have large constituent mass,
while the Wigner solution corresponds to DCS-unconfined phase (quark matter) and the corresponding mass function should take small values.
Since we have $A(p^{2})\sim 1$ for both Nambu and Wigner solutions,
it is easier to take the value of $B(p^{2})$ to identify the two solutions.
The calculated $B(p^{2}=0)$ as a function of quark chemical potential is shown in Fig.~\ref{fig:mu_B0}.

As can be seen from Fig.~\ref{fig:mu_B0},
   for every set of the truncation schemes,
   there is a coexistence region where both the Nambu and Wigner solutions exist.

The right boundary of this coexistence region is $\mu_{q}=308\;$MeV for all the three truncation schemes.
This is in accordance with the fact that we require that a singularity appears at that chemical potential.
After meeting the singularity, the DSE soon becomes unstable and the Nambu solution disappears.
This is in contradiction with the fact that there are hadron matter at the chemical potential larger than the critical one.
However, as we have stated in our previous work~\cite{Bai:2021non},
	the solution of the gap equation corresponds to a uniform and isotropic matter,
	while for hadron matter, the quarks are not uniformly distributed but are assembled as hadrons.
Therefore, the Nambu solution is only meaningful for $\mu_{q}<923/3\;$MeV since the vacuum is uniform.
For larger quark chemical potentials, we need to introduce the hadron degree of freedom when solving the DSE,
    or implement models to describe the hadron matter directly.

As for the Wigner solution, it still correctly corresponds to the deconfined quark matter,
	since in quark phase, the quarks are uniformly distributed.
The left boundary of the coexistence region is $\mu_{q}=276,\, 227$, $270\;$MeV for the DSE1, DSE2, DSE3 schemes, respectively.
We should mention that there are still Wigner solution below the chemical potential.
However, both the mass function and the chiral susceptibility ($\partial B/\partial m_{q}^{}$) are negative in small chemical potential region (see, e.g. Ref.~\cite{Qin:2010nq}),
and it has been proved that the negative chiral susceptibility corresponds to an unstable phase~\cite{Qin:2011dd,Gao:2016qkh}.
Therefore, we didn't plot those unstable Wigner solution in small chemical potential region.

\begin{figure}[!htb]
\includegraphics[width=0.43\textwidth]{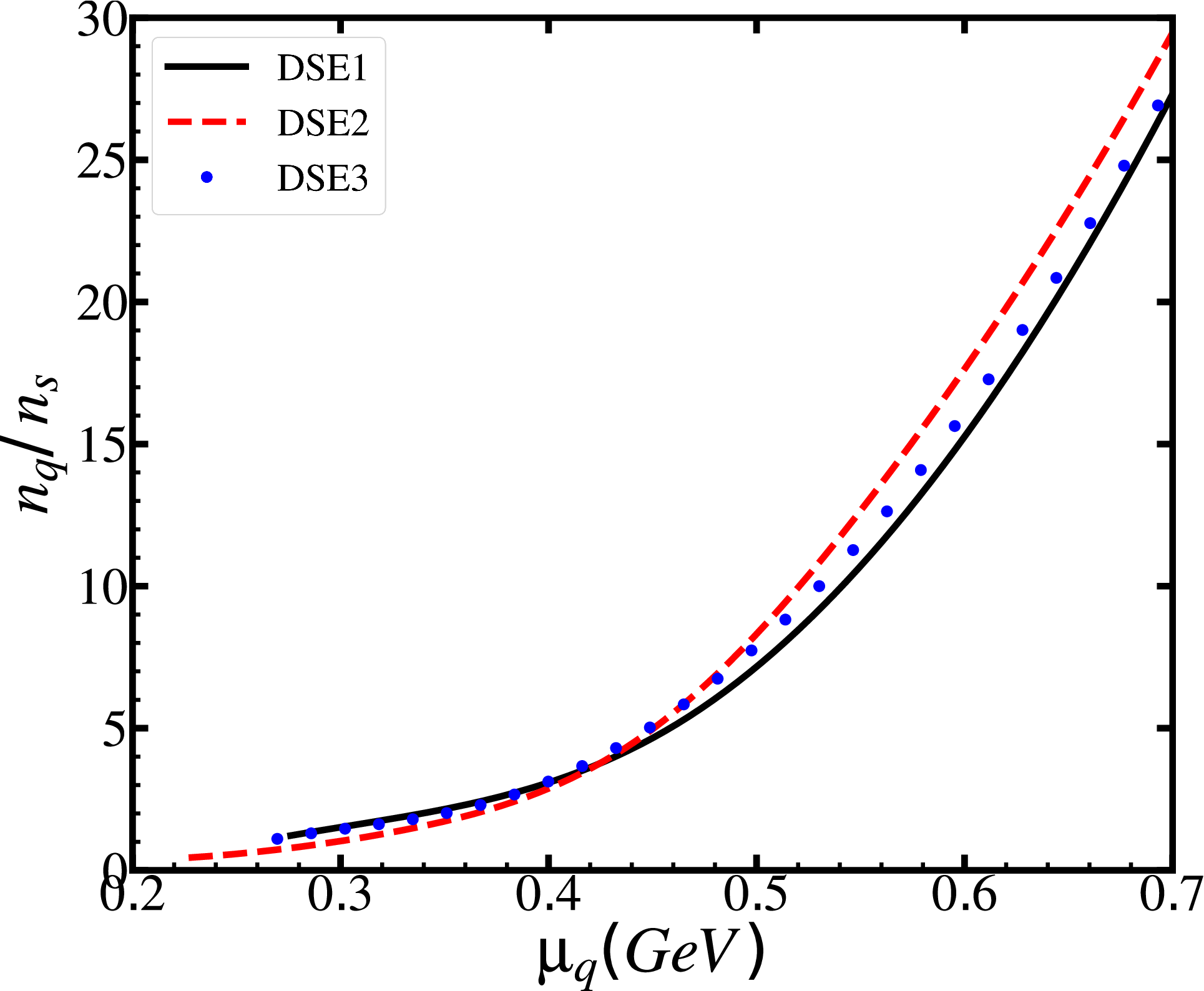}
\vspace*{-2mm}
\caption{(color online) The calculated $u$ and $d$ quark number density as a function of quark chemical potential in unit $n_{s}$, the saturation density of nuclear matter.
The black solid line corresponds to result with RL vertex and Gauss gluon model,
    the red dashed line corresponds to result with RL vertex and IC gluon model,
    and the blue dotted line corresponds to result with CLRQ vertex and IC gluon model.
}
\label{fig:mu_nq_ud}
\end{figure}

After having solved the gap equation,
	we can calculate the number density of quark matter with Eq.~(\ref{eqn:nq}).
The obtained $u$ and $d$ quark number density as a function of quark chemical potential for the three truncation schemes are shown in Fig.~\ref{fig:mu_nq_ud}.
From the figure, we can see that for different truncation schemes, the number density functions are not very different.

The pressure of a certain flavor of quark can be obtained by integrating the number density:
\begin{equation}\label{eqn:P_f}
P_{f}(\mu_{q})=\int_{\mu_{0}}^{\mu_{q}}\td\mu n_{q}(\mu_{q})+P_{f}(\mu_{0}),
\end{equation}
and the pressure of the quark matter is the sum of different flavors.
Mathematically, the starting of the integration $\mu_{0}$ can be any value as long as we know $P_{f}(\mu_{0})$ at that chemical potential
and the number density function between $\mu_{0}<\mu<\mu_{q}$.
In practice, we choose $\mu_{0}$ to be the left boundary of the coexistence region,
which is $\mu_{0}=276$, $227$, $270$MeV for the DSE1, DSE2, DSE3 model, respectively.
The value of $P_{f}(\mu_{0})$ can be regarded as a free parameter~\cite{Chen:2016ran},
but can also be studied in the framework of DSE.

In Ref.~\cite{Chen:2008zr},
the pressure difference between the Nambu and the Wigner solutions,
$\Delta P = P_{N}^{}-P_{W}^{}$ has been calculated using the ``steepest decent" approximation for $u$ and $d$ quarks.
The result is $\Delta P(\mu_{0})=4.19$, $5.36$, and $4.32\;\textrm{GeV}^{4}$ at $\mu_{0}=276$, $227$ and $270\;$MeV, respectively.
Since $P_{N}^{}$ corresponds to the pressure of the hadron phase,
it should be zero since the nuclear system is in vacuum at such a chemical potential.
Then, for $u$ and $d$ quarks,
we have $P_{u,d}(\mu_{0})=-\Delta P=-4.19,\, -5.36$, $-4.32\;\textrm{GeV}^{4}$ for the Wigner solution of the DSE1, DSE2, DSE3 truncation scheme, respectively.
The negative pressure means that the quark matter is unstable.
A bulk of the u/d quark matter will automatically shrink due to its negative pressure,
until the number density is high enough to have zero pressure and becomes metastable state.
The quark matter will becomes stable only after the hadron-quark phase transition, which we will study in the following.
For $s$ quark, we take $P_{s}(\mu_{0})=0$ as in previous work~\cite{Chen:2011my,Bai:2017wvk,Bai:2021non}.

As we have stated, the Nambu solution does not correspond to the real hadron matter after the emergence of matter due to the lack of hadron degree of freedom.
Therefore, in order to study the hadron-quark phase transition,
we implement phenomenological models for the hadron matter.
In this work, we adopt the relativistic mean field theory (RMF). 
For detail of this model, see Ref.~\cite{Typel:1999yq} or the appendix of our previous work~\cite{Bai:2021non}.

\begin{figure}[!htb]
\includegraphics[width=0.43\textwidth]{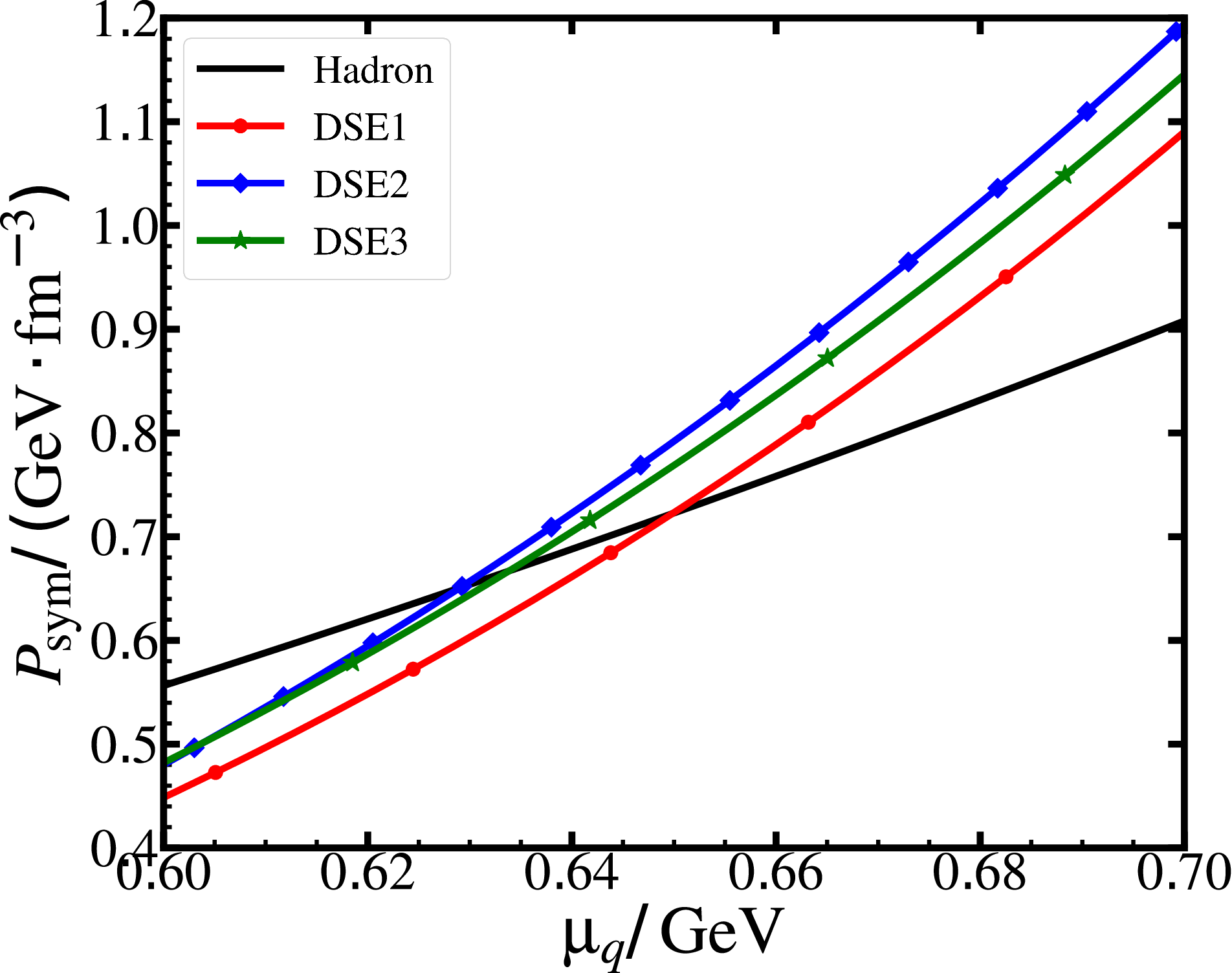}
\vspace*{-2mm}
\caption{(color online) The calculated pressure of the iso-symmetric matter as a function of quark chemical potential.
The black solid line corresponds to the result of hadron matter calculated using the relativistic mean field model,
    the red line with circle symbol corresponds to the result of quark matter with the RL vertex and Gauss gluon model,
    the blue line with square symbol corresponds to the result of quark matter with the RL vertex and IC gluon model,
    and the green line with star symbol corresponds to the result of quark matter with the CLRQ vertex and IC gluon model.
}
\label{fig:P_Mu_sym}
\end{figure}

In order for the phase transition to take place,
   the pressure and chemical potential should be the same in the two phases.
This corresponds to a cross point on the $P-\mu$ plane.
In Fig.~\ref{fig:P_Mu_sym},
   we present the calculated pressure as a function of quark chemical potential for hadron matter and quark matter with different sets of models.
For hadron matter, the chemical potential is $\mu_{n}=\mu_{p}=3\mu_{q}$ where $\mu_{n}$ and $\mu_{p}$ is the neutron and proton chemical potential, respectively,
    and for quark matter, we have $\mu_{u}=\mu_{d}=\mu_{q}$.

As can be seen from Fig.~\ref{fig:P_Mu_sym},
   the hadron line has cross points with all the quark lines in the three sets of quark models,
   and the chemical potential of the cross points correspond to the chemical potential of the chiral phase transition,
   which is $\mu_{q}=0.650$, $0.630$, $0.634\;$GeV for the DSE1, DSE2, DSE3 model, respectively.
This result show that the chemical potential corresponding to the chiral phase transition is not very different for different set of quark models,
i.e., the result converges with improved truncation schemes.
This can also be inferred from the fact that the number density function of the three models are almost the same.

\subsection{Phase transition and mix phase region for beta equilibrium matter}

\begin{figure}[!htb]
\includegraphics[width=0.45\textwidth]{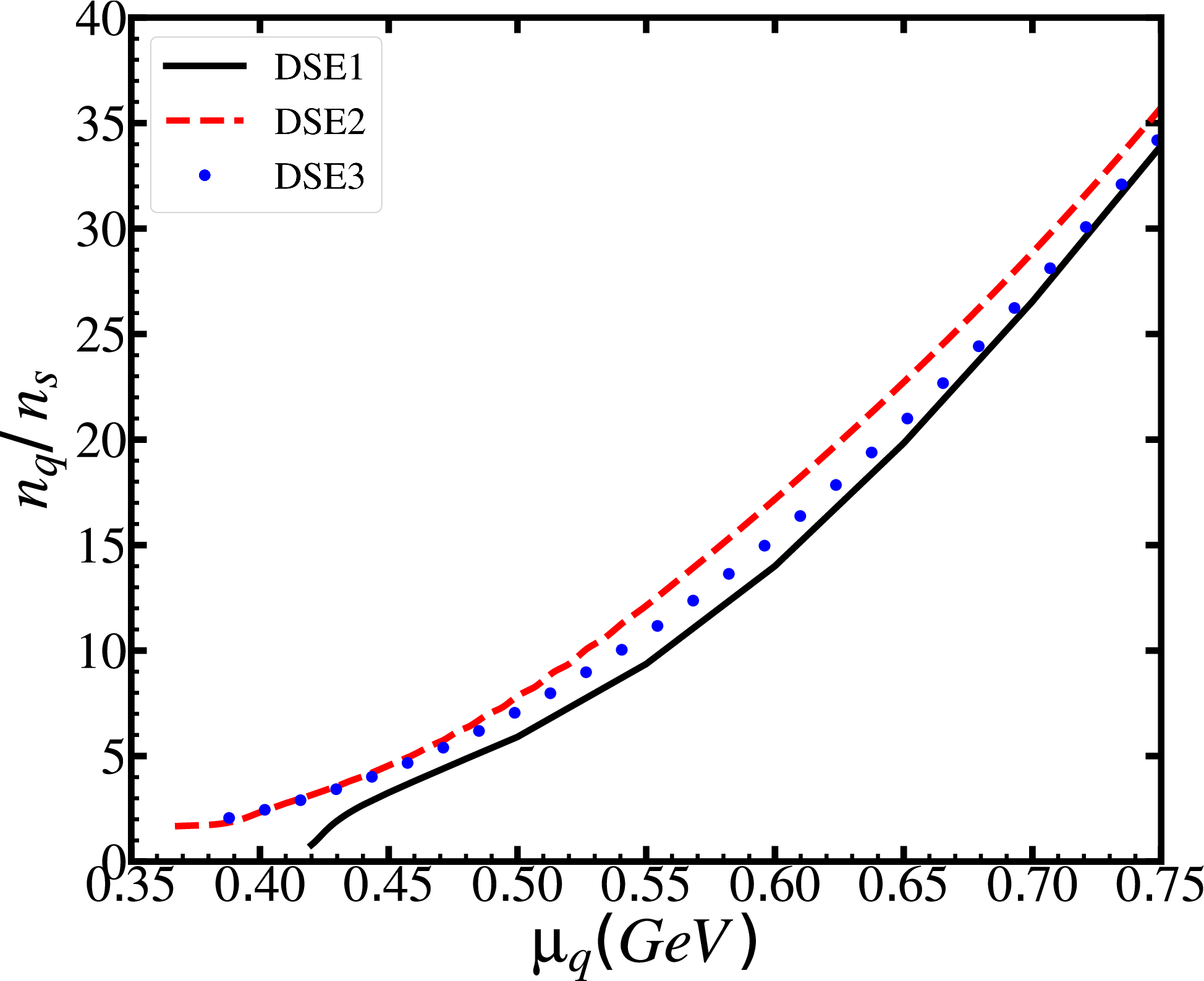}
\vspace*{-2mm}
\caption{ (color online) The calculated strange quark number density as a function of quark chemical potential in unit $n_{s}$, the saturation density of nuclear matter.
	The black solid line corresponds to the result using RL vertex and Gauss gluon model,
	the red dashed line corresponds to the result using RL vertex and IC gluon model,
	and the blue dotted line corresponds to the result using CLRQ vertex and IC gluon model.
}
\label{fig:mu_nq_s}
\end{figure}

In previous subsection, we have studied the phase transition of iso-symmetric matter.
However, for now we are unable to create the matter at such densities on earth,
	and we have to then take advantage of the astronomical observation of neutron stars for the theoretical study.
The astro-nuclear matter inside the neutron star is asymmetric, but in beta equilibrium and charge neutral.
Also, the possible appearance of strange quark and leptons (electron and muon) should be taken into account.
The obtained number density of strange quark as a function of quark chemical potential is shown in Fig.~\ref{fig:mu_nq_s}.
The chemical potential corresponding to the appearance of strange quark is $0.420$, $0.385$ and $0.388\;$GeV for DSE1, DSE2 and DSE3 set, respectively.

The pressure contribution of the strange quark can be calculated with Eq.~(\ref{eqn:P_f}).
In this paper, we assume that $P_{f}(\mu_{0})=0$ for strange quark.

The beta equilibrium condition requires that:
\begin{equation}\label{eqn:beta_q}
\begin{split}
& \mu_{d}^{}= \mu_{u}^{} + \mu_{e}^{} = \mu_{s}^{} \, ,\\
& \mu_{\mu^{-}}=\mu_{e}
\end{split}
\end{equation}
where $\mu_{e}$ is the chemical potential for electron, and $\mu_{\mu^{-}}$ is the chemical potential for muon.
The charge neutral condition requires that:
\begin{equation}\label{eqn:charge_neutral_q}
\frac{2n_{u}^{} - n_{d}^{} - n_{s}^{}}{3} - n_{e}^{} - n_{\mu^{-}}^{} = 0 \, ,
\end{equation}
where $n_{e}$ and $n_{\mu^{-}}$ is the number density for electron and muon, respectively.
In this paper, we assume that the leptons are free fermion gas,
   and their number density is:

\begin{equation}\label{eqn:nl}
\begin{split}
&n_{l}^{} = \frac{k_{Fl}^{3}}{3\pi^{2}} \, ,\\
&k_{Fl}^{2} = \mu_{l}^{2} - m_{l}^{2},
\end{split}
\end{equation}
where $l=e,\,\mu^{-}$.
In this work, we take $m_{e}=0.511\;$MeV and $m_{\mu^{-}}=105\;$MeV.

In the study of neutron star matter, it is easier to use baryon chemical potential $\mu_{B}$ instead of quark chemical potential,
   and their relation is:
\begin{equation}\label{eqn:muB_muq}
\mu_{B}=\mu_{u}+2\mu_{d}.
\end{equation}

Therefore, by combining Eqs.~(\ref{eqn:P_f}), (\ref{eqn:beta_q}), (\ref{eqn:charge_neutral_q}), (\ref{eqn:nl}) and (\ref{eqn:muB_muq}),
	for a given baryon chemical potential, we can calculate the pressure of the beta-equilibrium, charge neutral quark matter:
\begin{equation}
P_{Q}^{} = \sum_{f}P_{f}^{}(\mu_{f}) + P_{e}^{} + P_{\mu^{-}},
\end{equation}
and the energy density is:
\begin{equation}
\varepsilon_{Q} = n_{B}\mu_{B} - P_{Q}^{},
\end{equation}
where
\begin{equation}
n_{B}=\frac{n_{u}+n_{d}+n_{s}}{3}.
\end{equation}

There are several ways to describe the phase transition inside neutron stars.
The first is called ``Maxwell construction'',
and the phase transition condition is~\cite{Glendenning:2000}:
\begin{equation}
p_{H}^{}(\mu_{B})= p_{Q}^{}(\mu_{B}),
\end{equation}
where the subscripts $H$ and $Q$ denote the hadron, the quark sector, respectively.
The Maxwell construction is the most straight forward way of constructing the EOS of the matter involving the phase transition.

Another scheme is called ``Gibbs construction"~\cite{Glendenning:2000}.
Under this scheme, there is a mix phase region where both quark and hadron exist.
In the mixed region, the pressure of the two phases are the same.
The charge neutrality is not required separately in each phase,
but there exists a global charge neutral condition.
If we define the quark fraction $\chi$, with $0\le\chi\le 1$,
the phase transition condition can be expressed as:
\begin{equation}\label{eqn:condition1}
p_{H}^{}(\mu_{B},\mu_{e})= p_{Q}^{}(\mu_{B},\mu_{e}),
\end{equation}
\begin{equation}\label{eqn:condition2}
(1-\chi)n^{c}_{H}(\mu_{B},\mu_{e}) +\chi n^{c}_{Q}(\mu_{B},\mu_{e})=0,
\end{equation}
where $p_{H}^{}$ and $p_{Q}^{}$ is the pressure of the hadron, quark phase, respectively,
which are functions of both $\mu_{B}$ and $\mu_{e}$.
$n^{c}_{H}$ and $n^{c}_{Q}$ are the charge density of the two phases,
which can be determined by the corresponding hadron and quark model.

Then, combining Eqs.~(\ref{eqn:condition1}) and (\ref{eqn:condition2}),
together with the field equations of the two phases,
we can solve the $\mu_{B}^{}$ and $\mu_{e}$ with a given quark fraction $\chi$.
By taking $\chi=0$ and $1$,
we can calculate the left and right boundary of the mix phase region under charge neutral and beta equilibrium.

\begin{figure}[!htb]
\includegraphics[width=0.43\textwidth]{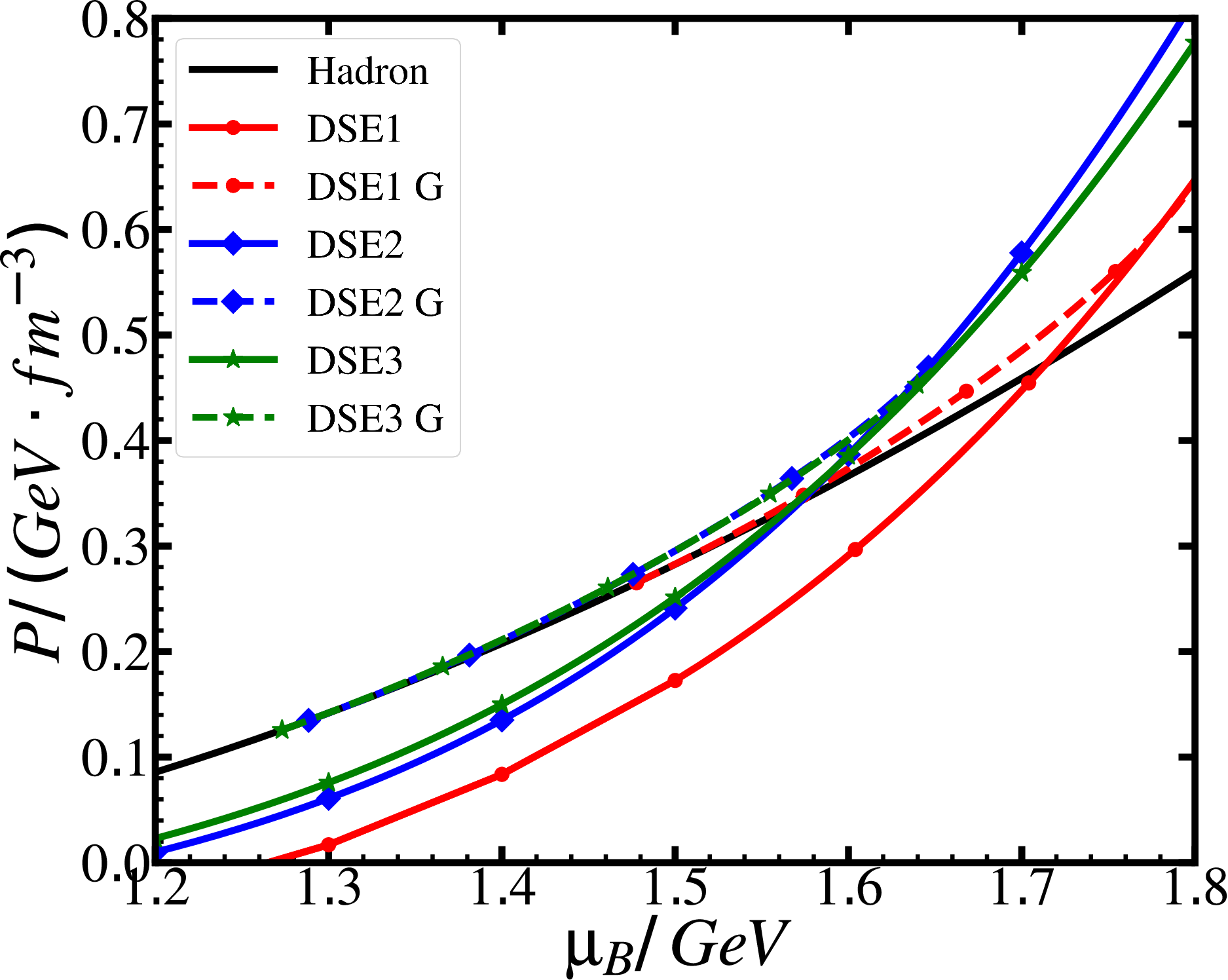}
\vspace*{-2mm}
\caption{(color online) The calculated pressure as a function of baryon chemical potential for charge neutral and beta equilibrium matter.
The black solid line corresponds to the pure hadron matter,
the solid lines with symbols corresponds to the pure quark matter,
and the dashed lines with symbols corresponds to the mix phase with the Gibbs construction,
and are denoted with a letter ``G" in the legend.
The red lines correspond to the result whose quark sector is calculated with the RL vertex and Gauss gluon model,
the blue lines correspond to the result whose quark sector is calculated with the RL vertex and IC gluon model,
and the green lines correspond to the result whose quark sector is calculated with the CLRQ vertex and IC gluon model.
}
\label{fig:P_Mu_beta}
\end{figure}

In Fig.~\ref{fig:P_Mu_beta},
we present the calculated pressure as a function of baryon chemical potential for neutron star matter.
In the figure, the cross points of black solid line with the colored solid lines correspond to the phase transition in the Maxwell construction,
   and the corresponding chemical potential is $\mu_{B,c}=1.71$, $1.57$ and $1.57\;$GeV for the DSE1, DSE2, DSE3 truncation scheme of the quark matter, respectively.
This is smaller than the phase transition chemical potential for iso-symmetric matter
(notice that there is a factor of 3 when comparing baryon chemical potential with quark chemical potential).
The colored dashed lines in Fig.~\ref{fig:P_Mu_beta} correspond to the mix phase in the Gibbs construction.
The two endpoints of a dashed line correspond to the boundary of the mix phase region,
and is
$1.48\le\mu_{B}\le 1.79$, $1.29\le\mu_{B}\le 1.65$ and $1.27\le\mu_{B}\le 1.66\;$GeV for the DSE1, DSE2, DSE3 truncation scheme, respectively.

For both Maxwell construction and Gibbs construction,
our results differs for different truncation schemes.
However, the discrepancy between DSE1 and DSE2 is much larger than that between DSE2 and DSE3.
This is the expected behavior,
since the result should gradually converge with the improvement of the truncation schemes,
and gets more and more reliable.
The small discrepancy between DSE2 and DSE3 is a strong hint that the result will not change much even if we further improve our truncation schemes.

The range of the mix phase region is different from the coexistence region we described in Sec.~\ref{sec::isosymmetric} and Fig.~\ref{fig:mu_B0},
where both Nambu and Wigner solutions are isosymmetric.
However, the mix phase region can provide some constraints on the Nambu and Wigner solution:
\begin{equation}\label{eqn:constraint_from_Gibbs}
	3 \mu_{W,c}^{} \le \mu_{G,1}^{} < \mu_{G,2}^{} \le 3\mu_{N,c}^{} \, ,
\end{equation}
where $\mu_{G,1}^{}$, $\mu_{G,2}^{}$ is the left and right boundary of the mix phase region in the Gibbs construction.
$\mu_{W,c}^{}$ is the critical chemical potential beyond which the Wigner solution has positive $B(p=0)$.
$\mu_{N,c}^{}$ is the upper limit for the Nambu solution to exist.
Eq.~(\ref{eqn:constraint_from_Gibbs}) means that, the region for the isosymmetric Nambu and Wigner solution to coexist,
should be larger than the region for the beta equilibrium hadron and quark phase to coexist,
otherwise the hadron or quark phase will correspond to a non-existing solution of the DSE.

From Fig.~\ref{fig:mu_B0} we can recognize easily that for the Wigner solution, $\mu_{W,c}^{}$ satisfies the first inequality in Eq.~(\ref{eqn:constraint_from_Gibbs}) for all different truncation schemes.
However, for Nambu solution, the last inequality in Eq.~(\ref{eqn:constraint_from_Gibbs}) is not satisfied.
As we have already pointed out in Sec.~\ref{sec::isosymmetric},
our Nambu solution is not reliable after the emergence of baryon number density,
since we have not taken into consideration the hadron degree of freedom.
Therefore, instead of directly take the result from our calculation,
we should implement the Eq.~(\ref{eqn:constraint_from_Gibbs}) as a constraint on the range of Nambu solution,
i.e., we have $\mu_{N,c}\ge 0.60$, $0.55$ and $0.55\;$GeV for DSE1, DSE2, DSE3 scheme, respectively.

\section{Equation of state and neutron star mass}\label{Sec::NeutronStar}

In order to get credible information for the cold dense hadron-quark phase transition,
on one hand, we need to improve our calculation with better and better truncation schemes,
on the other hand,
we have to take the compact star observations to check our theory since we are not able to create such dense matter on earth.

The most important observable of compact star is the maximum mass,
which is related to the EOS $P=P(\varepsilon)$ of the dense matter.

The energy density $\varepsilon$ and pressure $P$ of the dense matter under Maxwell construction is:
\begin{equation}
	\varepsilon_{\textrm{Maxwell}}^{} = \left\{ \begin{array}{ll}
		\varepsilon_{H}^{}, & \quad \textrm{if $\mu_{B}<\mu_{B,c}$}, \\[1mm]
		\varepsilon_{Q}^{}, & \quad \textrm{if $\mu_{B}>\mu_{B,c}$}; \\
	\end{array}\right.,
\end{equation}

\begin{equation}
	P_{\textrm{Maxwell}} = \left\{ \begin{array}{ll}
		P_{H}, & \quad \textrm{if $\mu_{B}<\mu_{B,c}$}, \\[1mm]
		P_{Q}, & \quad \textrm{if $\mu_{B}>\mu_{B,c}$}; \\
	\end{array}\right.,
\end{equation}
where $\varepsilon_{H}$ and $\varepsilon_{Q}$ is the energy density of the charge neutral hadron matter and the quark matter, respectively.
$P_{H}$ is the pressure of the hadron matter and $P_{Q}$ is the pressure of the quark matter.

For Gibbs construction, the EOS should be divided into three parts which are separately in the region:
$\mu_{B}\le \mu_{G,1}^{}$, $\mu_{G,1}^{} \le \mu_{B}\le \mu_{G,2}^{}$ and $\mu_{B}\ge\mu_{G,2}^{}$,
where $\mu_{G,1}^{}$, $\mu_{G,2}$ is the left, the right boundary of the mix phase region, respectively.

The energy density of the mixed phase consists of the contribution of the two phases.
\begin{equation}
\varepsilon_{M} = \chi \varepsilon_{Q}(\mu_B,\mu_e) + (1-\chi) \varepsilon_{H}(\mu_B,\mu_e),
\end{equation}
where the subscript $M$, $Q$ and $H$ correspond to the mixed, the quark and the hadron phase, respectively.
For a fixed $\mu_{B}$, we can solve $\mu_{e}$ from Eq.~(\ref{eqn:condition1}) and Eq.~(\ref{eqn:condition2}).
And the pressure of the mixed phase is:
\begin{equation}
P_{M}^{}(\mu_{B}) = P_{H}^{}\left(\mu_{B},\mu_{e}(\mu_{B})\right) = P_{Q}^{}\left(\mu_{B},\mu_{e}(\mu_{B})\right).
\end{equation}

The energy density and the pressure in the Gibbs construction are:
\begin{equation}
	\varepsilon_{\textrm{Gibbs}} = \left\{ \begin{array}{ll}
		\varepsilon_{H}, & \quad \textrm{if $\mu_{B}<\mu_{G,1}$}; \\[1mm]
             \varepsilon_{M}, & \quad \textrm{if $\mu_{G,1}\le\mu_{B}<\mu_{G,2}$}; \\[1mm]
		\varepsilon_{Q}, & \quad \textrm{if $\mu_{B}>\mu_{G,2}$}; \\
	\end{array}\right.
\end{equation}

\begin{equation}
	P_{\textrm{Gibbs}} = \left\{ \begin{array}{ll}
		P_{H}, & \quad \textrm{if $\mu_{B}<\mu_{G,1}$} ; \\[1mm]
        P_{M}, & \quad \textrm{if $\mu_{G,1}\le\mu_{B}<\mu_{G,2}$}; \\[1mm]
		P_{Q}, & \quad \textrm{if $\mu_{B}>\mu_{G,2}$}. \\
	\end{array}\right.
\end{equation}

\begin{figure}[!htb]
\includegraphics[width=0.43\textwidth]{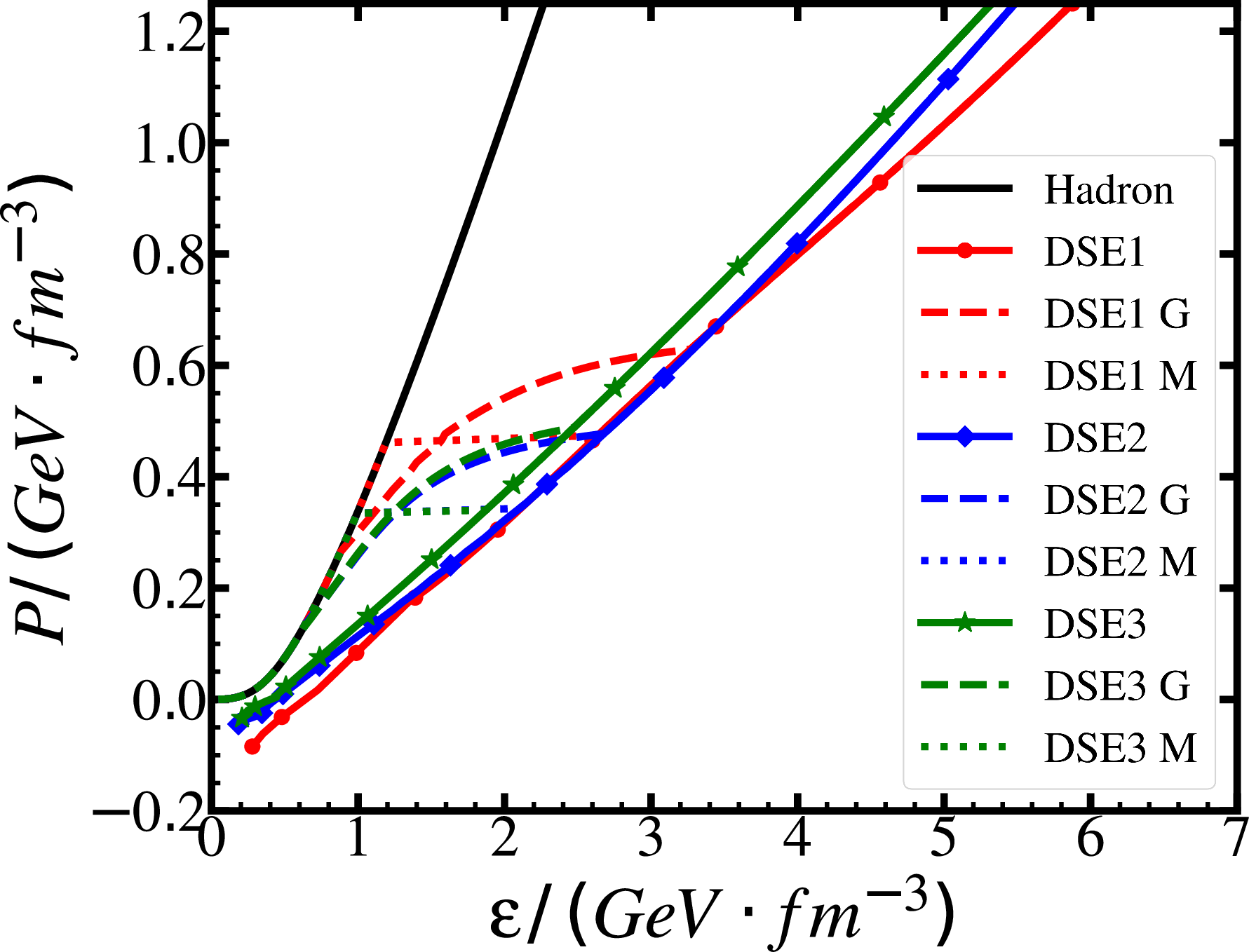}
\vspace*{-2mm}
\caption{(color online) Calculated EOS with the different models.
The black solid line corresponds to the pure hadron matter,
the colored solid lines with symbols correspond to the pure quark matter.
The colored dashed lines correspond to the mixed phase in the Gibbs construction,
and are denoted with a letter ``G" in the legend.
The colored dotted lines correspond to the Maxwell construction,
and are denoted with a letter ``M" in the legend.
The red lines correspond to the results with the RL vertex and Gauss gluon model,
the blue lines correspond to those with the RL vertex and IC gluon model,
and the green lines correspond to those with the CLRQ vertex and IC gluon model.
}
\label{fig:EoS}
\end{figure}

The calculated EOS of the pure hadron, the pure quark and the hybrid matter is shown in Fig.~\ref{fig:EoS}.
The colored solid lines with symbols correspond to the results of the pure quark matter calculated with different truncation schemes.
As we can see from the figure,
the EOS of the quark matter is softer than that of the hadron matter, no matter what truncation we take.
Although the three quark lines are not very different,
the EOS of the DSE1 truncation scheme is relatively softer than that of the DSE2 and DSE3 truncation scheme.
The colored dashed lines correspond to the EOS of mixed phase with Gibbs construction.
Unlike the pure quark EOS, the mixed phase with DSE1 quark sector is the most stiff one.
The horizontal dotted line is the EOS with the Maxwell construction.
However, in compact stars, in order to resist the gravity,
	the EOS must be increasing, and the horizontal region of the EOS will not appear in the compact star.

We mention that, in the middle density region,
the EOSs of DSE2 and DSE3 are almost identical,
which means that our result converges with the improvement of truncation schemes.
However, at extremely large density, the discrepancy between DSE2 and DSE3 gradually increases.
This means that we need to apply even better truncation schemes at that density.
In our current paper, we are only interested in the density ranges corresponding to the hadron-quark phase transition,
and we would like to neglect the discrepancy beyond those ranges.

The mass-radius relation of neutron stars can be calculated
by solving the Tolman-Oppenheimer-Volkov (TOV) equation:
\begin{equation}
\frac{\textrm{d}P}{\textrm{d}r}=-\frac{G}{r^{2}}(m + 4\pi Pr^{3})(\varepsilon + P)\left(1-2\frac{Gm}{r}\right)^{-1} \, ,
\end{equation}
where $G$ is the gravitational constant and $m=m(r)$ is the mass inside radius $r$:
\begin{equation}
m(r)=\int_{0}^{r} 4\pi R^{2} \varepsilon(R) \textrm{d}R.
\end{equation}

Then given the EOS as input,
and with a given center density,
we can integrate the TOV equation from inside out to get the mass and radius of the neutron star.

For pure hadron star and hybrid star, at small density region, we take the Baym-Pethick-Sutherland (BPS) EOS~\cite{Baym:1971pw}.
For pure quark star, we integrate to the surface where the pressure is zero.

\begin{figure}[!htb]
\includegraphics[width=0.43\textwidth]{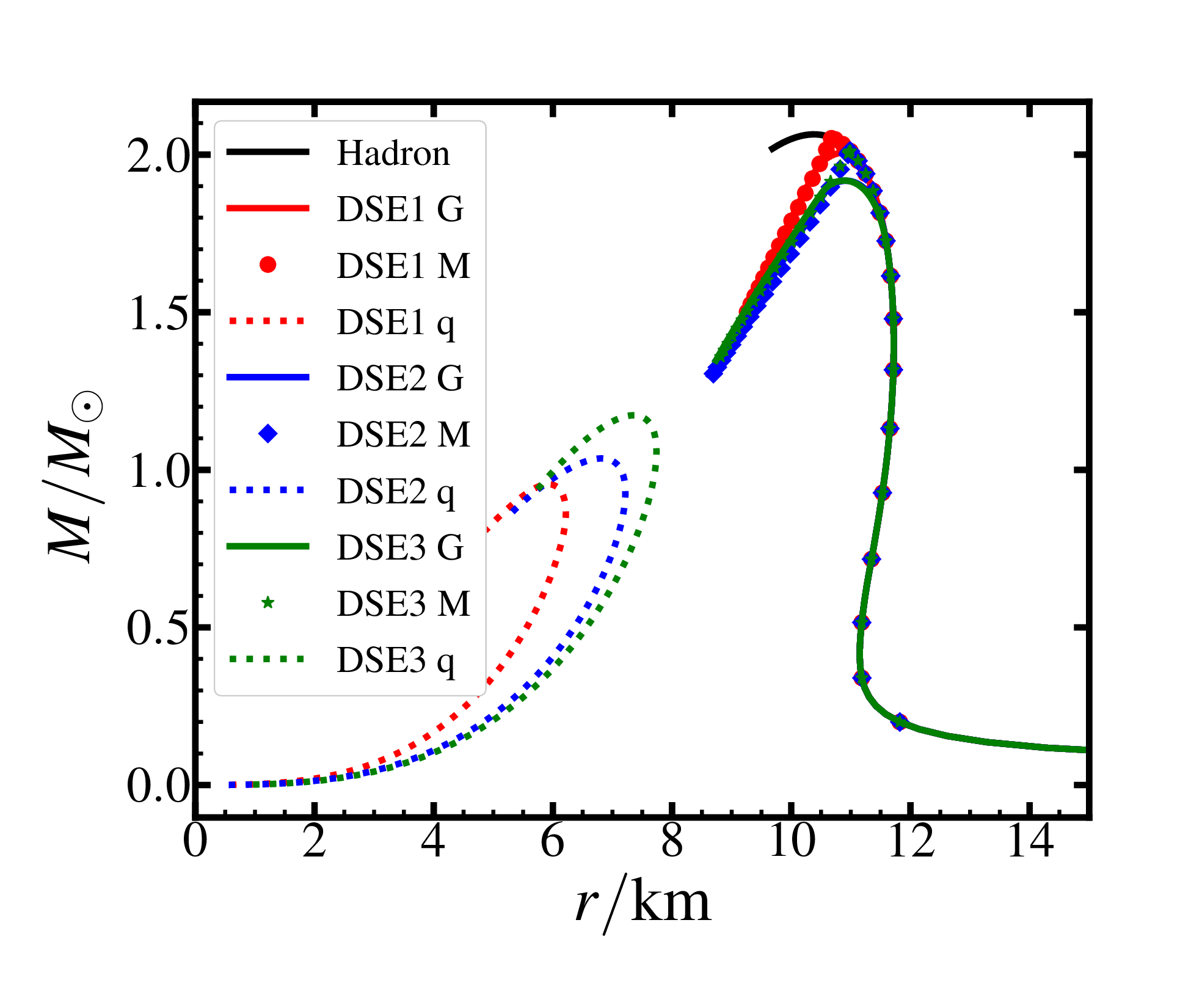}
\vspace*{-2mm}
\caption{(color online) Calculated mass-radius relation of the compact stars.
The black solid line corresponds to the result of pure hadron star.
The colored solid lines correspond to the result of hybrid star with Gibbs construction,
and are denoted with a letter ``G" in the legend.
The colored symbols correspond to the result of hybrid star with Maxwell construction,
and are denoted with a letter ``M" in the legend.
The colored dotted lines correspond to the result of pure quark stars,
and are denoted with a letter ``q" in the legend.
The red lines and symbols correspond to those with the RL vertex and Gauss gluon model,
the blue lines and symbols correspond to those with the RL vertex and IC gluon model,
and the green lines and symbols correspond to those with the CLRQ vertex and IC gluon model.
}
\label{fig:m-r}
\end{figure}

The calculated mass-radius relation with the three models are shown in Fig.~\ref{fig:m-r}.
It is apparent that the maximum mass of the pure hadron star is $2.06\, M_{\odot}$.
For pure quark star, however, the maximum mass is $0.95$, $1.03$, $1.17\,M_{\odot}$ for the DSE1, DSE2, DSE3 truncation scheme, respectively.
This result is in accordance with the stiffness of the EOS for the quark sector.

As for the hybrid star, we see that the mass-radius curves of the DSE2 and DSE3 sectors are almost the same,
   no matter for Gibbs construction or Maxwell construction.
This can be deduced from the fact that the EOS of the hybrid star matter with DSE2 and DSE3 models are almost the same at small energy densities (see Fig.~\ref{fig:EoS}).

The maximum mass for hybrid star with Maxwell construction is $2.05$, $2.01$, $2.01M_{\odot}$ for the DSE1, DSE2, DSE3 truncation scheme, respectively.
For all these quark truncation schemes, the mass of hybrid star will reduce after the phase transition occurs.
However, if the mass of the hybrid star is smaller for larger central density,
	the star will be unstable against oscillations.
Therefore, there will not be a quark core inside the hybrid star in the Maxwell construction.

The maximum mass for hybrid star with Gibbs construction is $2.00$, $1.92$, $1.92M_{\odot}$ for the DSE1, DSE2, DSE3 truncation scheme, respectively.
Using Gibbs construction, the hybrid star reaches maximum mass after the appearance of the quark core.
When implementing the DSE2 and DSE3 truncation schemes, the maximum mass is smaller than $2M_{\odot}$, even without the inclusion of hyperons.

As we have shown in previous sections,
the results from DSE2 and DSE3 have already shown good convergence,
and are not likely to change much if we continue to improve our truncation schemes.
Therefore, in order to get a heavier hybrid star, 
we should use better hadron models or construction schemes, instead of modifying the quark sector.

For example, we can take more realistic model for the hadron sector, e.g., the Brueckner-Hartree-Fock (RBHF) theory~\cite{Baldo:2007wm,Shen:2019dls,Qin:2023zrf}.
We can also apply 3-window construction~\cite{Masuda:2012ed,Masuda:2012kf,Kojo:2015fua},
which have already been used to construct a heavy enough hybrid star~\cite{Bai:2017wvk,Qin:2023zrf}.

Apart from the mass of neutron stars, 
another important astronomical observation is the gravitational wave (GW) from binary neutron star merger
~\cite{Demorest:2010bx,Antoniadis:2013pzd,Fonseca:2016tux,NANOGrav:2017wvv,NANOGrav:2019jur,Linares:2018ppq}.
By analyzing the GW data, it is believed that the tidal deformability $\Lambda_{1.4}$ of a $1.4M_{\odot}$ neutron star can be constrained to a narrow range
~\cite{Annala:2017llu,LIGOScientific:2018hze,Coughlin:2018miv,Malik:2018zcf,LIGOScientific:2018cki}.
However, in our calculation, the hadron-quark phase transition will not occur in neutron stars with mass $1.4M_{\odot}$,
no matter what truncation schemes we use.
Therefore, the tidal deformability does not provide constraint on our DSE study in this work.
We mention that, for pure hadron star,
the RMF model we use will give correct $\Lambda_{1.4}$ which satisfies astronomical data, see e.g. Ref.~\cite{Xia:2023omv}.

\section{Summary and Remarks}\label{Sec::summary}

In this paper, we make use of the first-principle Dyson-Schwinger equation approach to study the cold dense QCD matter.
We solve the gap equation with different truncation schemes and compared the results.
We have considered three combinations of the gluon and the interaction vertex:
RL vertex with Gaussian gluon, RL vertex with infrared-constant gluon, and CLRQ vertex with infrared-constant gluon.
The calculated result is expected to gradually converge with better gluon and vertex used.

By solving the gap equation at a series values of the chemical potential,
we find that hadron matter will stay in the DCSB vacuum of QCD until a very high critical chemical potential.
This chemical potential is in contradiction with the nuclear liquid-gas phase transition chemical potential,
at which the hadron matter is expected to appear.
Also, by improving the truncation scheme,
the critical chemical potential does not converge to a fixed value.
Therefore, we proved that it is necessary to modify the truncation schemes, especially in high chemical potential region.

We introduced a modification factor into the truncation schemes, and fixed the parameter by requiring that the quark number density becomes non-zero at the nuclear liquid-gas phase transition chemical potential.

After the modification, the solutions with different truncation schemes converge
automatically at three different chemical potentials:
the vacuum, where the original truncation schemes reproduce the meson properties;
the liquid-gas critical chemical potential, where matter begins to emerge from vacuum;
and extremely large chemical potential, where QCD matter approaches the asymptotic freedom.

We then make use of the modified truncation schemes to study the hadron-quark phase transition,
and find that the results also converge, especially for RL+IC and CLRQ+IC.

We take the Wigner solution to describe the quark phase,
and implement the relativistic mean field model to describe the hadron phase,
and then analyze the phase transition chemical potential.
The calculation shows that for the different truncation schemes we take,
the phase transition chemical potential are nearly the same for iso-symmetric matter,
which is $0.650$, $0.630$ and $0.634\;$GeV for the RL+Gauss, RL+IC and CLRQ+IC truncation, respectively.

For beta equilibrium and charge neutral matter which is relevant with those in neutron stars,
the phase transition baryon chemical potential under Maxwell construction is $\mu_{B}=1.71$, $1.57$ and $1.57\;$GeV,
and the mixed phase region is $1.48\le\mu_{B}\le 1.79$, $1.29\le\mu_{B}\le 1.65$ and $1.27\le\mu_{B}\le 1.66\;$GeV
for the RL+Gaussian, RL+IC and CLRQ+IC truncation, respectively.
For the simplest truncation, RL+Gaussian, the phase transition chemical potential is larger than that of the other two sets of truncation.
And for the improved schemes, although there are still difference, the phase transition takes place at almost the same chemical potential,
which is a good proof of the convergence.

The same argument can also be applied to the EOS and the mass-radius relation of the hybrid star.
For the RL+IC and CLRQ+IC schemes, the EOS are very close to each other, especially in case of small energy density.
And the mass-radius relations of the hybrid star are almost the same for these two schemes.

The obtained results show that,
    after introducing a chemical potential dependent modification factor to the coupling strength,
    and requiring that the matter begins to appear at $\mu_{q,c}=923/3\;$MeV,
    different vertex and gluon models give similar results on the phase transition and EOS for the cold dense matter,
    especially when we take the improved gluon and vertex.
This proves that our choice of modification scheme is reliable.

However, our work can still be improved further.
Firstly, although we have required that the matter appears at the nuclear liquid-gas phase transition chemical potential,
we have not yet correctly reproduced the first-order nature of this phase transition automatically.
This is due to the lack of hadron degree of freedom,
and might be amended by considering the back reaction effect of hadrons on the quark propagator~\cite{Eichmann:2015kfa}.

Also, after applying the improved vertex and gluon model,
	the calculated maximum mass of hybrid stars are smaller than $2M_\odot$.
This is not likely to change much by improving the quark sector,
and should be amended by implementing the 3-window interpolation rather than Gibbs or Maxwell construction~\cite{Masuda:2012ed,Masuda:2012kf,Kojo:2015fua,Bai:2017wvk,Qin:2023zrf},
or use other hadron models such as Brueckner-Hartree-Fock (RBHF) theory~\cite{Baldo:2007wm,Shen:2019dls,Qin:2023zrf}.

We should also mention that the possible appearance of color-superconducting phase~\cite{Barrois:1977xd,Alford:2001dt,Alford:2007xm} might change our result.
For example, it is argued that the hadron-quark phase transition might become a crossover~\cite{Brandes:2021pti,Baym:2017whm,Fukushima:2020cmk}.
There have already been studies about color-superconducting with DSE~\cite{Muller:2013pya,Muller:2016fdr},
but this method has not yet been taken to study the properties of cold dense neutron star matter.
We will further explore this possibility in our future work.

\section*{Acknowledgement}
This work was supported by the National Natural Science Foundation of China under Grant No. 12175007, No. 12205353 and No. 12247107, 
and also the China Postdoctoral Science Foundation under Grant No. 2022M723230,
CAS Project for Young Scientists in Basic Research (YSBR060).

\appendix


%

\end{document}